\documentclass[revtex4,iop]{emulateapj}

\begin{document}
\title{Laboratory and On-Sky Validation of the Shaped Pupil Coronagraph's Sensitivity to Low-Order Aberrations With Active Wavefront Control}
\author{
Thayne Currie\altaffilmark{1,2,3},
N. Jeremy Kasdin\altaffilmark{4},
Tyler D. Groff\altaffilmark{5},
Julien Lozi\altaffilmark{1},
Nemanja Jovanovic\altaffilmark{1,6},
Olivier Guyon\altaffilmark{1,7,8,9},
Timothy Brandt\altaffilmark{10,11},
Frantz Martinache\altaffilmark{12},
Jeffrey Chilcote\altaffilmark{13,14},
Nour Skaf\altaffilmark{1,15},
Jonas Kuhn\altaffilmark{16},
Prashant Pathak\altaffilmark{1},
Tomoyuki Kudo\altaffilmark{1}
 }
\altaffiltext{1}{National Astronomical Observatory of Japan, Subaru Telescope, National Institutes of Natural Sciences, Hilo, HI 96720, USA}
\altaffiltext{2}{NASA-Ames Research Center, Moffett Field, California 94035}
\altaffiltext{3}{Eureka Scientific, 2452 Delmer Street Suite 100. Oakland, CA 94602-3017}
\altaffiltext{4}{Department of Mechanical and Aerospace Engineering, Princeton University, Princeton, NJ 08544, USA}
\altaffiltext{5}{NASA-Goddard Space Flight Center, Greenbelt, MD, USA}
\altaffiltext{6}{Department of Astronomy, California Institute of Technology, Pasadena, CA, USA}
\altaffiltext{7}{Astrobiology Center, National Institutes of Natural Sciences, 2-21-1 Osawa, Mitaka, Tokyo, Japan}
\altaffiltext{8}{Steward Observatory, University of Arizona, Tucson, 933 N. Cherry Avenue, Tucson, AZ 85721, USA}
\altaffiltext{9}{College of Optical Science, University of Arizona, 1630 E. University Boulevard, Tucson, AZ 85719, USA}
\altaffiltext{10}{Department of Physics, University of California-Santa Barbara, Santa Barbara, CA, USA}
\altaffiltext{11}{Astrophysics Department, Institute for Advanced Study, Princeton, NJ, USA}
\altaffiltext{12}{Observatoire de la Cote d'Azur, Boulevard de l'Observatoire, Nice, 06304, France}
\altaffiltext{13}{Department of Physics, Stanford University, Palo Alto, CA, USA}
\altaffiltext{14}{Department of Physics, University of Notre Dame, 225 Nieuwland Science Hall, Notre Dame, IN, 46556, USA}
\altaffiltext{15}{Institut d'Optique Graduate School Paris XI, 2 Av. Augustin Fresnel Palaiseau, 91120 France}
\altaffiltext{16}{Institute for Particle Physics and Astrophysics, ETH Zurich, Wolfgang-Pauli-Strasse 27, 8093 Zurich, Switzerland}
\begin{abstract}
We present early laboratory simulations and extensive on-sky tests validating of the performance of a shaped pupil coronagraph (SPC) behind an extreme-AO corrected beam of the \textit{Subaru Coronagraphic Extreme Adaptive Optics} (SCExAO) system.    In tests with the SCExAO internal source/wavefront error simulator, the normalized intensity profile for the SPC degrades more slowly than for the Lyot coronagraph as low-order aberrations reduce the Strehl ratio from extremely high values (S.R. $\sim$ 0.93--0.99) to those characteristic of current ground-based extreme AO systems (S.R. $\sim$ 0.74--0.93) and then slightly lower values down to S.R. $\sim$ 0.57.
On-sky SCExAO data taken with the SPC and other coronagraphs for brown dwarf/planet-hosting stars HD 1160 and HR 8799 provide further evidence for the SPC's robustness to low-order aberrations.  From H-band Strehl ratios of 80\% to 70\%, the Lyot coronagraph's performance vs. that of the SPC may degrade even faster on sky than is seen in our internal source simulations.   The 5-$\sigma$ contrast also degrades faster (by a factor of two) for the Lyot than the SPC.   The SPC we use was designed as a technology demonstrator only, with a contrast floor, throughput, and outer working angle poorly matched for SCExAO's current AO performance and poorly tuned for imaging the HR 8799 planets.  Nevertheless, we detect HR 8799 cde with SCExAO/CHARIS using the SPC in broadband mode, where the S/N for planet e is within 30\% of that obtained using the vortex coronagraph.  The shaped-pupil coronagraph is a promising design demonstrated to be  robust in the presence of low-order aberrations and may be well-suited for future ground and space-based direct imaging observations, especially those focused on follow-up exoplanet characterization and technology demonstration of deep contrast within well-defined regions of the image plane.

\end{abstract}
\keywords{planetary systems, stars: early-type, stars: individual: HR 8799} 
\section{Introduction}
Over the past decade, large ground-based telescopes assisted with facility adaptive optics systems have revealed the first direct images of young, self-luminous superjovian planets around nearby stars
\citep[e.g.][]{Marois2008,Lagrange2010,Carson2013,Rameau2013,Currie2014}.   Dedicated  \textit{extreme} adaptive optics (extreme AO) systems coupled with coronagraphy have now improved our ability to image young planets and candidates closer to their host stars and added to our inventory \citep[e.g.][]{Macintosh2015,Currie2015a, Milli2017, Chauvin2017}.  However, directly detecting mature solar system analogues in reflected light ($\lesssim$ 10$^{-8}$ contrast) from the ground or space requires advances in wavefront control and coronagraphy to suppress stellar halo light with far greater precision.

In contrast to the traditional Lyot coronagraph with an opaque mask which blocks light in the focal plane, several recently-developed, promising coronagraph designs instead achieve deep contrasts by apodizing and remapping the pupil plane transmission profile
\citep{Guyon2003,Kasdin2003,Soummer2005}.   One of these designs is the \underline{\textit{shaped pupil coronagraph}} \citep[SPC;][]{Kasdin2003,Kasdin2005,Kasdin2007}, which uses a binary transmission function at the pupil plane to reshape the point-spread function in the image plane from an Airy pattern to one with concentrated much higher-contrast, ``discovery zone" regions more suitable for imaging faint exoplanets.   
Lacking a focal plane mask, the original design for the SPC (considered in this work) may be far less sensitive to low- order aberrations intrinsic to the optical path and those induced by ground or space-based telescope vibrations, in comparison to coronagraph designs including a focal plane mask \citep{Green2004,Blackwood2013}.   More recent SPC designs, such as the one selected as baseline architecture for WFIRST-CGI now include a focal plane mask.  Recent laboratory tests suggest that the SPC with a focal plane mask, when used behind a well-corrected wavefront, can achieve raw contrasts below 10$^{-8}$ in narrowband optical light from space \citep{Cady2016}.

Despite its success in highly controlled environments, there are no published studies of the SPC used with a real wavefront control system behind a real telescope looking at a real star to begin to assess its performance in practice.   Other advanced coronagraphs, typically those operating in the image plane, have been validated from ground-based telescopes, used in combination with near-infrared focused  extreme AO systems or with the best facility AO systems at longer wavelengths \citep[e.g.][]{Beuzit2008,Macintosh2014,Serabyn2017,Kuhn2017}.   Compared to their raw, expected performance, the achieved contrasts from these systems have been substantially degraded due to residual wavefront errors, especially low-order aberrations.   One example is the high-performance vector vortex coronagraph on Keck and Subaru/SCExAO, whose peak rejection is up to a factor of five worse than predicted in the absence of aberrations \citep{Mawet2005,Serabyn2017,Kuhn2017}.   Robust laboratory simulations of the SPC's performance against progressively larger residual halo light and drops in Strehl ratio plus on-sky observations of planet-hosting stars with an extreme AO system coupled with the SPC can 1) verify the insensitivity of the SPC's raw contrast to aberrations versus those of other coronagraphs and 2) validate the SPC's capability to image faint planets at small angular separations.

In this study, we report the first validation of the shaped-pupil coronagraph behind a well-corrected wavefront but in the presence of low-order aberrations, using laboratory and on-sky observations of the SPC with the Subaru Coronagraphic Extreme Adaptive Optics (SCExAO) project  at the 8.2 $m$ Subaru Telescope on Maunakea \citep{Jovanovic2015} coupled with the CHARIS integral field spectrograph \citep{Peters2012,Groff2015,Groff2016}.  Recent results show that SCExAO has achieved  extreme AO capability and, when combined with both standard and advanced image-plane coronagraphs (e.g. the Lyot and vector vortex coronagraphs), yields contrasts rivaling those of other  extreme AO systems \citep[][]{Currie2017}.    While its performance right now is largely limited by low-order aberrations, especially tip-tilt induced by encoders on the telescope drive system, this weakness makes SCExAO an ideal testbed for validating the performance of the SPC in the laboratory and on-sky, looking at real astrophysical objects.

\begin{figure}
\centering
\includegraphics[scale=0.75]{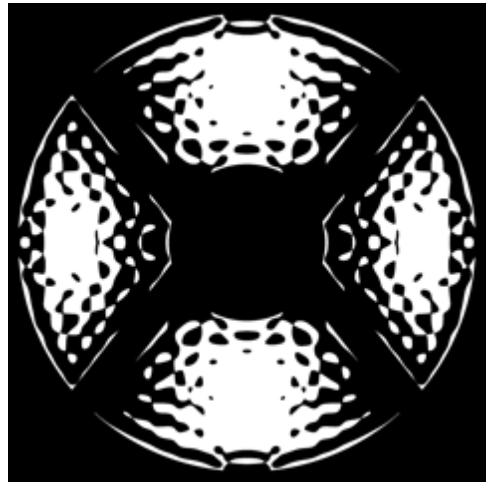}
\caption{The shaped-pupil coronagraph design used in this study.}
\label{spcmaskapp}
\end{figure}

\section{The Shaped Pupil Coronagraph for SCExAO}
The shaped pupil coronagraph (SPC) we tested behind SCExAO was designed using the Micro/Nano Fabrication Laboratory at Princeton University and is first depicted in Figure 6 of \citet{Carlotti2012}.   The mask was fabricated by patterning the chrome onto a 3.2 mm thick fused silica substrate with a 10\arcsec{} wedge to mitigate ghosts.   The substrate was polished to a $\lambda$/10 root-mean squared (rms) to minimize residual wavefront error. Due to fabrication constraints the substrate is not anti-reflective coated.  This design has approximately 50\% effective pupil area. The central obstruction, outer diameter, and spiders are all padded by 2\% to account for alignment error, providing a 3 degree clocking tolerance on the registration of the coronagraph mask to the incident telescope pupil. 

\begin{figure*}[ht]
\centering
\includegraphics[scale=0.49,trim=0mm 0mm 0mm 10mm,clip]{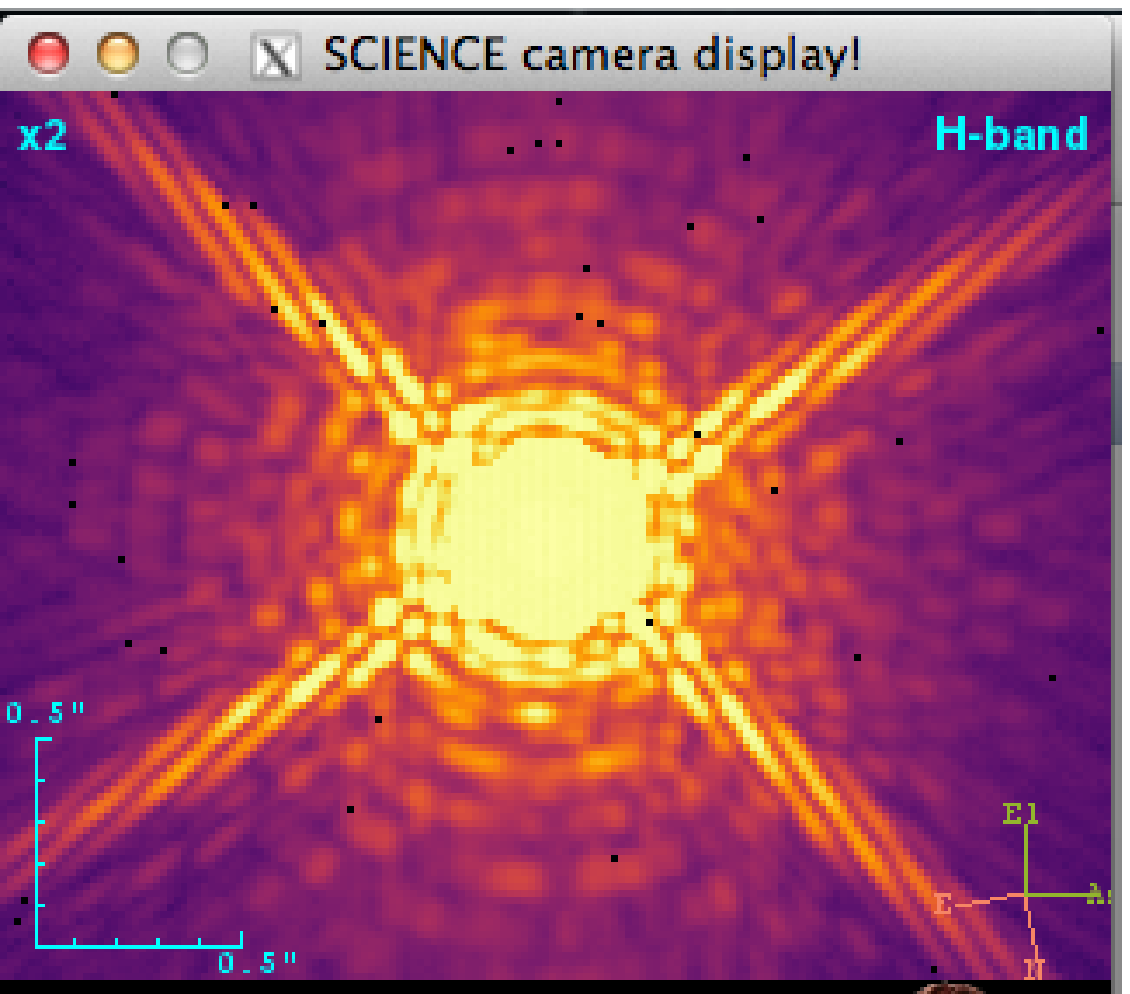}
\includegraphics[scale=0.49,trim=0mm 0mm 210mm 10mm,clip]{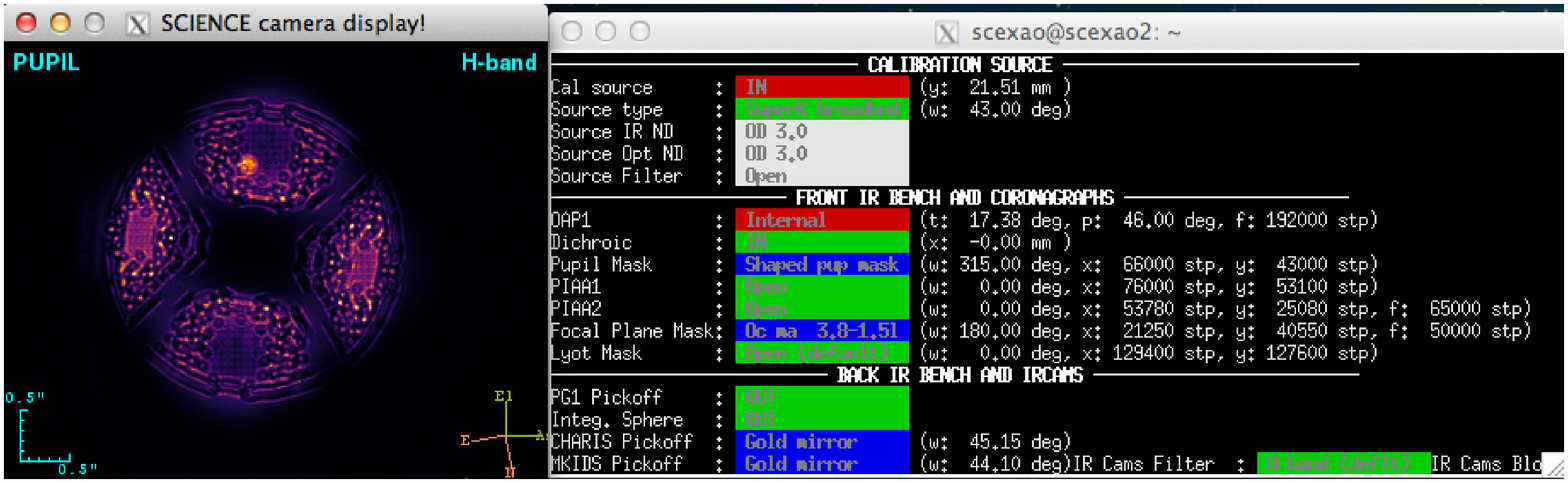}
\includegraphics[scale=0.49,trim=0mm 0mm 210mm 10mm,clip]{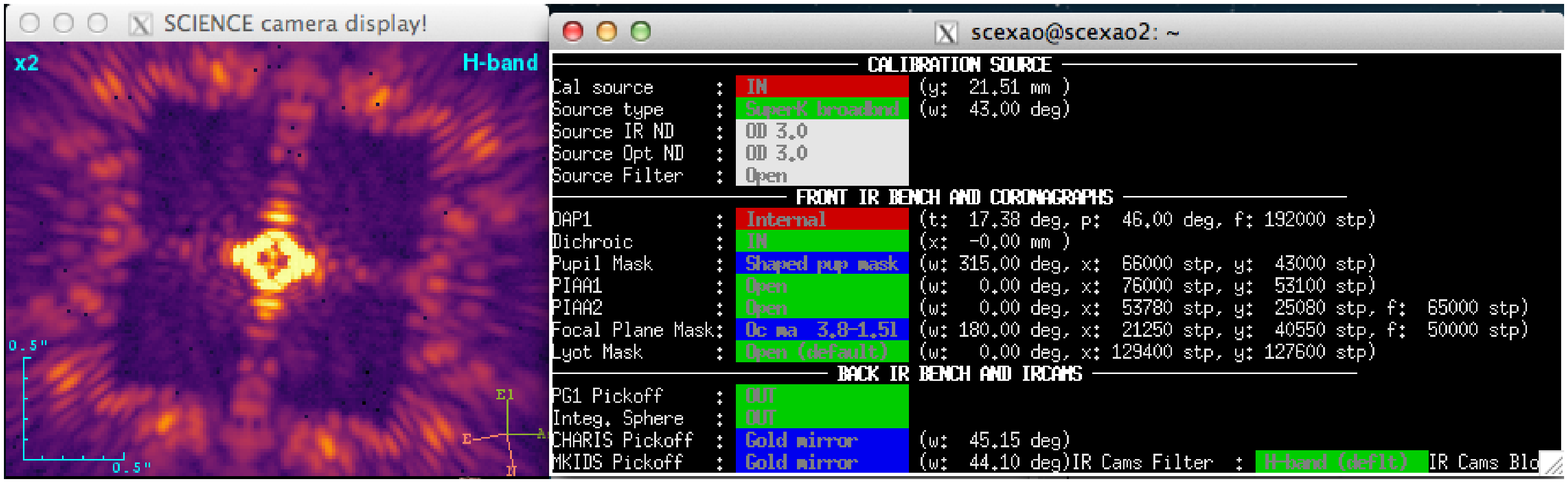}
\vspace{-0.02in}
\caption{\textit{(left)} Unocculted SCExAO point-spread function, (\textit{middle}) pupil plane image of SCExAO+the shaped pupil coronagraph, and  (\textit{right}) focal plane image of SCExAO behind the shaped pupil coronagraph with a 3.8 $\lambda$/D (0\farcs{}16 at 1.6 $\mu m$) focal plane mask (used to prevent saturation) after precise optical alignment.}
\label{spc_pupfoc}
\end{figure*}

  The mask is shown in Figure \ref{spcmaskapp}; the appearance of the Subaru/SCExAO pupil and image plane in the lab with this pupil mask placed in is shown in Figure \ref{spc_pupfoc}.   The pupil image shows the varying, ``ripple" transmission profile characteristic of shaped pupils.  

 In the image plane, the mask yields a four-quadrant dark hole demarcated by bright diagonal regions with an inner working angle of 3.5 $\lambda$/D and an outer working angle of 16 $\lambda$/D, which is roughly 0\farcs{}15 and 0\farcs{}67 for Subaru/SCExAO at 1.6 $\mu m$, respectively. 
    The SPC was originally designed in 2012 to yield a null depth appropriate for Subaru's facility adaptive optics capabilities at the time, which achieved $H$-band Strehl ratios of $\sim$ 30--40\% in good conditions and contrasts of $\sim$ 10$^{-5}$ at 0\farcs{}75 and 1.0\arcsec{} over 30 minute-long sequences using advanced observing and image processing techniques \citep[e.g.][]{Brandt2014}.   In comparison, with the vector vortex and standard Lyot coronagraphs SCExAO can achieve contrasts a factor of 10 to 100 deeper from 1\arcsec{} to 0.25\arcsec{} \citep[][T. Brandt and T. Currie 2017, unpublished]{Currie2017}.
 
 Therefore, since this SPC is not optimized for SCExAO science, our analysis focuses on the \textit{relative} performance of the SPC under different conditions and compared to the relative performance of other coronagraphs used with SCExAO.
 
 \begin{deluxetable*}{llcrcccccccc}
\setlength{\tabcolsep}{0pt}
\tablecolumns{8}
\tablecaption{SCExAO Coronagraph Simulations}
\tiny
\tablehead{ {WFE (Inputed, Effective)}&{Low-Order Coeff.}
& {Strehl Ratio} & {} & {Log(Norm. Intensity) (Lyot, SPC)} & {}\\
 {(nm)}& {} & {(at 1.55 $\mu m$)} & {(0\farcs{}2-- 0\farcs{3})} &{(0\farcs{45}--0\farcs{55})}}
\startdata
0, 0 & 0 & 0.99 &  -4.11, -3.82&     -4.98, -4.82\\
35.9, 49.1 & 0.12 & 0.96 & -3.93, -3.61&  -4.85, -4.73\\
 43.4, 68.4 & 0.14 & 0.93 & -3.85, -3.55 & -4.77, -4.70\\ 
 50.0, 85.2 & 0.15 & 0.89 & -3.78, -3.52 & -4.72, -4.66\\
 67.0, 118.9 & 0.18 & 0.79 & -3.61, -3.38 & -4.60, -4.57\\
75.0, 134.7 & 0.20 & 0.74 & -3.54, -3.34 & -4.54, -4.53\\
100.0, 183.8 & 0.25 & 0.57 & -3.37, -3.18 & -4.38, -4.41\\
 \enddata
 \tablecomments{The ``inputed" wavefront error is the value entered in to the SCExAO turbulence simulator.   The ``effective" wavefront error is the wavefront error inferred given the measured Strehl ratio according to Mar\'echal's formula \citep{Marechal1947}.   The normalized intensities listed are the median-averaged halo intensity (not rms) divided by the stellar flux within one FWHM (roughly 40 mas).}
\label{simdata}
\end{deluxetable*}

 \section{Performance of the Shaped Pupil Coronagraph with the SCExAO Turbulence Simulator}
 \subsection{Experimental Set-Up}
This work utilized the SCExAO instrument which has been described in detail in \citet{Jovanovic2015}.  Here we only highlight the key SCExAO features needed to understand its usage in these experimental tests. 
 Briefly, we modeled the performance of the shaped-pupil coronagraph using the SCExAO internal light source and wavefront error simulator.   The internal source emitted light centered on 1550 nm with a bandwidth of 50 nm (3\% effective bandwidth) comparable to the spacing of spectral channels with SCExAO coupled with the CHARIS integral field spectrograph in low-resolution, broadband mode.   The plate scale of the internal SCExAO camera is 0\farcs{}15/pixel: also comparable to CHARIS.  To establish a reference point for the SPC's halo suppression, we also obtained images with a standard Lyot coronagraph.  In both cases, we used an occulting mask with a diameter of 217 mas.   Lyot coronagraph data utilized a Lyot stop, while SPC data were obtained without one\footnote{The occulting spot was used with the SPC to prevent saturation.}.  
 
To model the performance of each coronagraph design in the presence of low-order aberrations, we used the simulator to write (to file) a wavefront error map (Table \ref{simdata}).  The map is defined by the input simulator wavefront error and ``low-order coefficient", which parameterizes the fidelity of the correction for the lowest spatial frequencies.   The connection between the inputed simulation wavefront error and the actual wavefront error/Strehl ratio was determined empirically using unocculted images.  We adopt a wind speed of 4.0 m s$^{-1}$ to define the rate at which the static wavefront is dragged across the SCExAO pupil.  The $H$-band Strehl ratios for which we obtained data cover the theoretical limit assuming no wavefront errors (S.R. $>$ 0.99) to those slightly better than achieved under excellent conditions with facility AO systems on Keck and Subaru (S.R. $\sim$ 0.57) \citep{vanDam2004}.   In between these two extremes, we focus on Strehl ratios that may be achieved with a future ground-based extreme AO system (S.R. $\sim$ 0.96), those that are characteristic of SCExAO right now (T. Currie 2017, unpublished) for bright stars observed under very good seeing conditions (S.R. $\sim$ 0.89--0.93), and those for fainter stars or observations under slightly poorer conditions (S.R. $\sim$ 0.74--0.79). 

   Each sequence of data taken with the SPC or Lyot coronagraph consisted of several thousand co-added exposures with individual exposure times of 0.5--5 ms, the exact value tuned to ensure that the detector response was linear, and added up to a cumulative exposure time comparable to that for a typical on-sky CHARIS data cube ($t$ $\sim$ 30-60 $s$).   Integration times for the SPC were about a factor of 5-10 longer than for the Lyot because of the SPC's significantly lower throughput (see \S 4.1.2).  Satellite spots produced by modulating SCExAO's deformable mirror provide absolute flux calibration \citep{Jovanovic2015b} with a contrast ratio of $\Delta$m = 6.412 $\pm$ 0.050 at 1.55 $\mu m$.  
To assess each coronagraph's performance at each Strehl ratio and value for the low-order coefficient is the \textit{normalized intensity profile} (NIP): the median-averaged intensity at each separation divided by the stellar flux within one FWHM ($\sim$ 0\farcs{}04).

\subsection{Methodology for Estimating the Sensitivity of the Shaped-Pupil Coronagraph to Low-Order Aberrations}
Precisely quantifying the sensitivity of different coronagraphs to low-order aberrations is challenging, as it requires disentangling the coronagraphs' sensitivities to low-order aberrations from differences in their intrinsic performance or that set by the telescope aperture.   Thus, in this section we explain and justify our approach in detail. 

 \citet{Mazoyer2017} provides a useful reference point for how we approach this issue.  They used a simple model to assess the tip-tilt sensitivity of a system with a fixed coronagraph design \citep[from ][]{Fogarty2017} but two different apertures: that of WFIRST-CGI and a segmented aperture.   As shown in their Figure 7, the system performance vs. tip-tilt within a fixed dark hole region (3--10 $\lambda$/D) for both apertures has multiple regimes: a flat slope at the lowest tip-tilt values (tip-tilt $\lesssim$ 10$^{-4}$ $\lambda$/D) where the performance is limited by the choice of aperture and is independent of tip-tilt, a rising and constant slope at tip-tilt amplitudes greater than 10$^{-3}$ $\lambda$/D where the performance is limited by the coronagraph and is sensitive to tip-tilt, and a transition region at intermediate tip-tilt amplitudes.  In the coronagraph performance-limited regime, naturally the slopes are identical and the two coronagraphs are equally sensitive to tip-tilt since in fact the exact same coronagraph design is used.   
 
 Thus, if different coronagraphs (e.g. Lyot, SPC) are used with the same aperture, the design less sensitive to tip-tilt and other low-order aberrations at a given angular separation should have a shallower slope in $\Delta$(NIP)/$\Delta$(SR) provided that low-order aberrations are what degrade the quality of the point-spread function, as is the case with our simulations.
 
 This result for a given angular separation can be generalized to all separations of interest as follows.   First, for each coronagraph we divide the normalized intensity profile (NIP) at each Strehl ratio by the profile at a Strehl ratio of 0.99 (NIP$_{SPC, Lyot}$(S.R. $<$ 0.99)/NIP$_{SPC, Lyot}$(S.R. = 0.99)), to assess how the profile for a given coronagraph degrades with lower Strehl ratios driven by larger low-order aberrations.  Next, we divide NIP at each Strehl ratio for the Lyot by the corresponding NIPs for the SPC.  We call this ratio the \textit{relative degradation in performance} or RDP.  
 
 If the shaped pupil is less sensitive to low-order aberrations than the Lyot coronagraph over a range of angular separations, then a) RDP is greater than 1 and b) RDP becomes progressively larger at lower Strehl ratios.   The second condition is required to differentiate between a coronagraph that is more sensitive to low-order aberrations versus one that is simply has a better performance at very high Strehl\footnote{Note that this approach is consistent with the proper interpretation of Figure 7 in \citet{Mazoyer2017}.   There, we would first normalize the contrast for each aperture by its value in the limit of no tip-tilt.  Then we would divide the normalized contrast for the WFIRST aperture (blue curve) by the normalized contrast for the segmented aperture (labeled SCDA; red curve).   Doing so would give RDP $>$ 1 because the SCDA contrast floor (used for normalization) is 100 times deeper.
  However, beyond a tip-tilt value larger than 10$^{-3}$ $\lambda$/D, RDP is constant (roughly a factor of 50).   Thus, the second condition is not satisfied, and there is no difference in coronagraphic tip-tilt sensitivity in the two cases.}.

   \begin{figure}[h]
\centering
\includegraphics[scale=0.45,trim=0mm 0mm 0mm 0mm,clip]{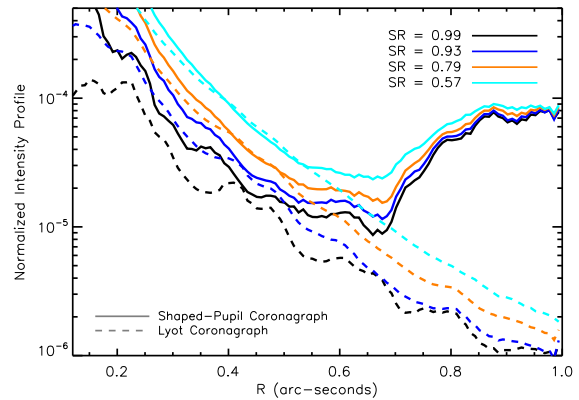}
\vspace{-0.02in}
\caption{Normalized intensity profile (residual halo intensity/star signal within one full-width at half-maximum) profile obtained for SCExAO internal source data with the shaped pupil (solid lines) and Lyot coronagraphs (dashed lines) shown for Strehl Ratios between 0.99 and 0.57.    The rise in contrast for the shaped pupil at $r$ $>$ 0\farcs{}75 is due to the SPC's transmission profile which redistributes off-axis light  at small angular separations out to larger separations.}
\label{lyotspc_simcontrast}
\end{figure}

Note that there are key differences between our experiments and the simulations from \citet{Mazoyer2017} that may make for a more direct test of the robustness of different coronagraphs to low-order aberrations.  \citet{Mazoyer2017} model a highly stable space-borne environment.  For a substantial amount of their explored phase space,  low-order aberrations such as tip-tilt are so small ($<$ 10$^{-3}$ $\lambda$/D) that the aperture, not the coronagraph, limits the system's performance.   Tip-tilt alone induced during our simulations is much larger and plainly visible in individual frames.  

Furthermore, ground-based extreme AO systems currently in operation are nowhere as stable as space-borne platforms; 
  Tip-tilt alone often measures in excess of 0.1 $\lambda$/D on SCExAO even at very high Strehl ratios; SPHERE's dedicated mirror corrects tip-tilt only up to 0.05 $\lambda$/D at H band \citep{Fusco2015}.   Simulations with the vortex coronagraph behind SCExAO show that low-order terms substantially degrade performance even at Strehl ratios of 90--95\% \citep{Kuhn2017}.   Generally speaking, even at Strehl ratios in excess of 80\% low-order terms are plausibly the dominant stellar leakage term (see Appendices A2--A7 in \citealt{Kuhn2017}).  Only after additional steps like the implementation of a coronagraphic low-order wavefront sensor will low-order aberrations be substantially better mitigated: e.g. reducing tip-tilt down to 10$^{-3}$ $\lambda$/D \citep{Singh2017}.   
\subsection{Results}

 Figure \ref{lyotspc_simcontrast} displays the normalized intensity profiles for the SPC and Lyot coronagraphs under the assumption of no wavefront error and a range of wavefront errors induced by low-order aberrations.  Coupled with the Lyot coronagraph, the residual stellar halo's contrast is $\sim$ 10$^{-4}$, 10$^{-5}$, and 2$\times$ 10$^{-6}$ at 0\farcs{}25, 0\farcs{}5, and 0\farcs{}75.   These values are intermediate between the contrast of the unobscured Subaru pupil and what would likely be achieved with a higher performance coronagraph such as an APLC \citep[e.g. see Figure 4 in ][]{Soummer2005}.  Despite its much lower throughput and more relaxed contrast floor, the SPC's NIP in absence of wavefront errors is only about a factor of two poorer at $\rho$ $\sim$ 0\farcs{}2--0\farcs{}55 before becoming substantially poorer exterior to $\sim$ 0\farcs{}75\footnote{The bright diagonal regions for the SPC subtend a small fraction of the halo exterior to $r$ $\sim$ 0\farcs{}2--0\farcs{}3 and have a very minor impact on our calculated profiles.}.   
 
 \begin{figure}[ht]
\vspace{-0.1in}
\centering
\includegraphics[scale=0.475,trim=5mm 0mm 0mm 0mm,clip]{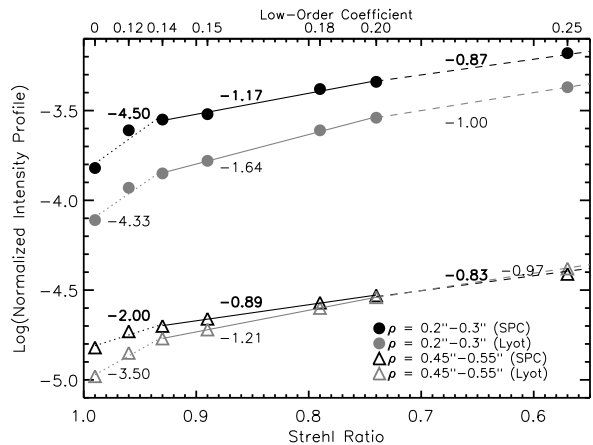}
\vspace{-0.025in}
\caption{Testing the SPC's sensitivity to low-order aberrations with SCExAO laboratory data, showing the log(contrast) vs. low-order coefficient and Strehl ratio for the shaped pupil (black, bold symbols) and Lyot (light gray symbols) at two representative angular separations: $\rho$ = 0\farcs{}2--0\farcs{}3 (triangles) and 0\farcs{}45--0\farcs{}55 (circles).    The best-fit linear trend in three regimes -- extremely high Strehl (0.93--0.99, dotted lines), very high Strehl characteristic of ground-based extreme AO systems (0.74--0.93, solid lines), and lower Strehl (0.57--0.74, dashed lines) -- are shown for both coronagraphs.   The slopes for these linear fits are listed as bold, black letters for the SPC and light gray letters for the Lyot.    With the possible exception of the highest Strehl limit at 0\farcs{}2--0\farcs{}3, the SPC always has a shallower slope, suggestive of a weaker sensitivity to low-order aberrations.
}
\label{lyotspc_simrelv0}
\end{figure}

\begin{figure}[ht]
\includegraphics[scale=0.48,trim=0mm 0mm 0mm 0mm,clip]{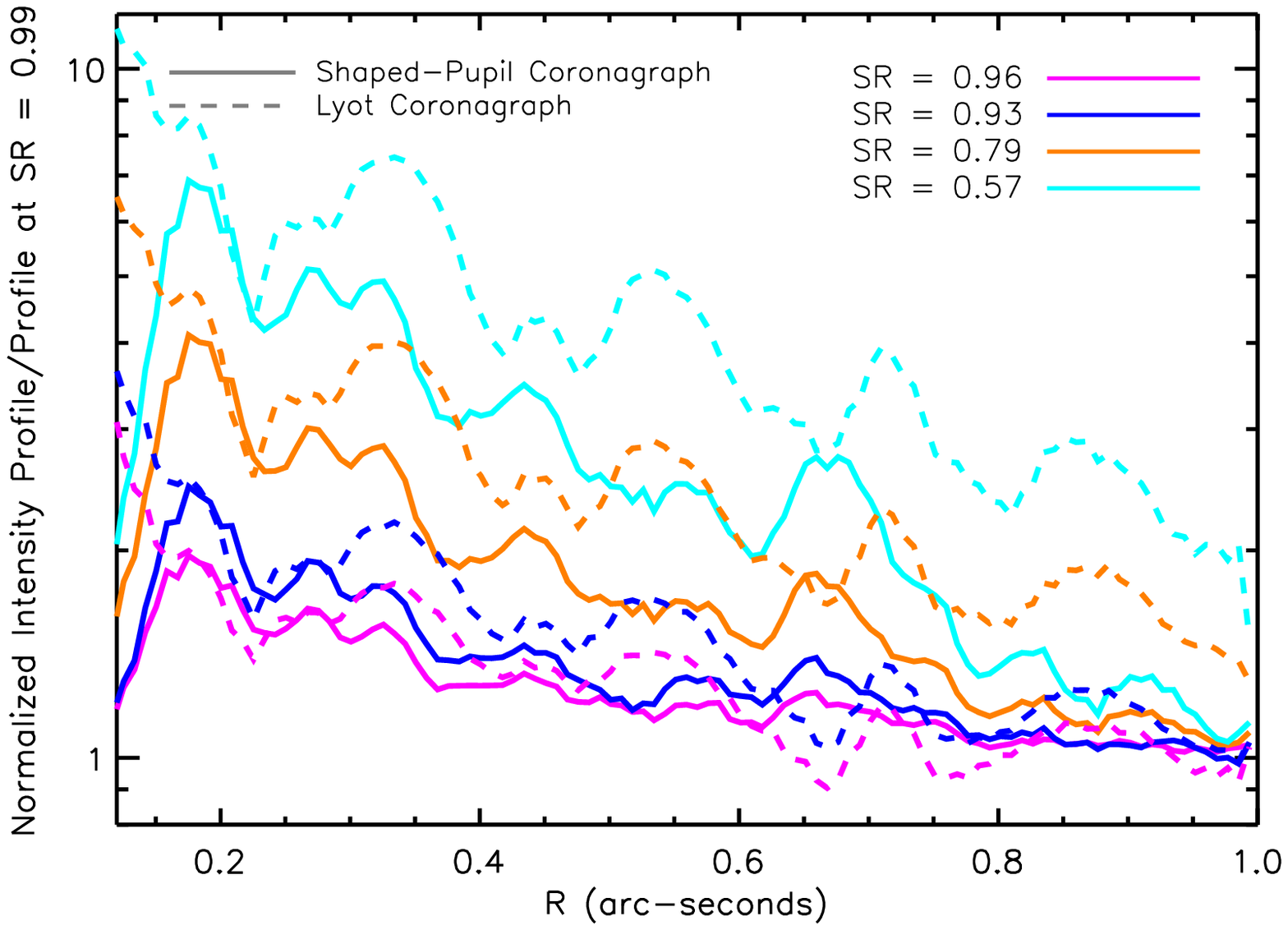}\\
\includegraphics[scale=0.48,trim=0mm 0mm 0mm 0mm,clip]{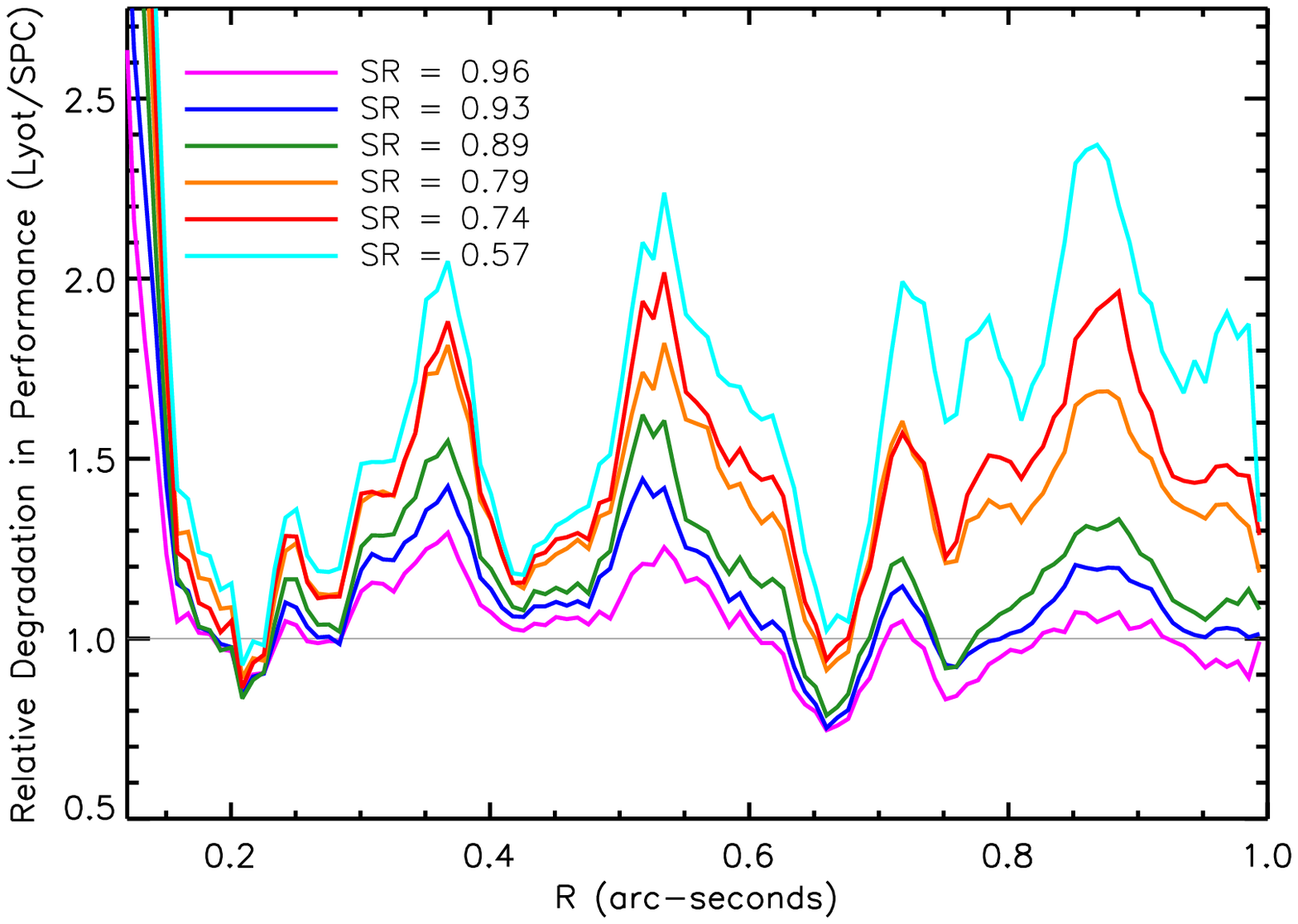}
\vspace{-0.02in}
\caption{  \textit{(Top)} The laboratory-derived normalized intensity profile at a given Strehl relative to its value in absence of aberrations (SR $>$ 0.99) for the shaped pupil and Lyot coronagraphs.   (\textit{Bottom}) The relative degradation in performance (RDP) between the Lyot and shaped-pupil coronagraphs.
Larger values for lower Strehl ratios means that the Lyot coronagraph's performance is degrading more as low-order aberrations increase.    With few exceptions,  the SPC's performance degrades more slowly than the Lyot  over the angular separations covering the SPC's dark hole ($r$ $\sim$ 0\farcs{}2--0\farcs{}6), where the designs' performance in absence of low-order aberrations is comparable, suggestive of a weaker sensitivity to low-order aberrations.
}
\label{lyotspc_simrel}
\end{figure}

Figure \ref{lyotspc_simrelv0} displays and the righthand columns of Table \ref{simdata} list the normalized intensity vs. low-order coefficient (LOC) and Strehl ratio for the Lyot and SPC at two representative angular separations: $\rho$ = 0\farcs{}2--0\farcs{}3 and $\rho$ = 0\farcs{}45--0\farcs{}55.   These curves reveal three regimes: a fast drop in performance at the highest Strehl ratios (S.R. $\sim$ 0.99 to 0.96 or 0.93), a slightly slower decay in regimes covering typical extreme AO performance on SCExAO/SPHERE/GPI (S.R. $\sim$ 0.93 to 0.74) and a very slow decay down to Strehl ratios characteristic of the best conventional AO systems (S.R. $\sim$ 0.74 to 0.57).  

At Strehl ratios relevant for dedicated exoplanet imaging systems currently in operation (S.R. $\sim$ 0.74--0.93, LOC = 0.14--0.20), the SPC is less sensitive to low-order aberrations at $\rho$ = 0\farcs{}2--0\farcs{}3 and 0\farcs{45}--0\farcs{}55.    Its log(NIP) decays roughy as dlog(NIP)/d(SR) $\sim$ -1.17 and $\sim$ -0.89 at $\rho$ = 0\farcs{}2--0\farcs{}3 and $\rho$ = 0\farcs{}45--0\farcs{}55, respectively.  In comparison, the Lyot's performance decays as dlog(NIP)/d(SR) $\sim$ -1.64 and $\sim$ -1.21 at $\rho$ = 0\farcs{}2--0\farcs{}3 and $\rho$ = 0\farcs{}45--0\farcs{}55.   At lower Strehl ratios (S.R. 0.74 to 0.57), the shaped pupil's performance also decays more slowly than the Lyot\footnote{We find the same qualitative results if instead we try to determine the slope of the entire range of S.R. = 0.93 to 0.57 with one line.   The slope for the SPC is shallower, indicating a reduced sensitivity to low-order aberrations.}.  At the very highest Strehl ratios, the situation for our two ranges of angular separations is somewhat ambiguous, as the SPC's performance decays far more slowly than the Lyot at $\rho$ = 0\farcs{}45--0\farcs{}55 but slightly faster at 0\farcs{}2--0\farcs{}3, although Table \ref{simdata} shows that quantitatively the difference between the drop in performance between the Lyot and SPC in the latter case is extremely small.


Figure \ref{lyotspc_simrel} provides a way of visualizing the sensitivity of different coronagraph designs to low-order aberrations for the full range of angular separations of interest, first by normalizing the NIP by its value at a Strehl ratio of 0.99 (top panel).   Going from a Strehl of 0.99 to 0.96, the contrasts for both coronagraphs degrade by a factor of $\sim$ 1.25 to 2 from $\rho$ $\sim$ 0\farcs{}75 down to 0\farcs{}25.  However,  between 0\farcs{}2 to 0\farcs{}6, the SPC's NIP brightens 10\% less.   

The bottom panel of Figure \ref{lyotspc_simrel} plots the RDP vs. Strehl ratio and is a generalization of the results in Figure \ref{lyotspc_simrelv0} and Table \ref{simdata} for the full range of angular separations of interest.    If the SPC is less sensitive to low-order aberrations, its performance decay rate will be shallower: curves for progressively lower Strehl ratios will lie at larger y axis values.  This is exactly what we see.  
 For extreme AO corrections in the near-infrared (S.R. $\sim$ 0.74--0.93), the Lyot's contrast at 0\farcs{}25--0\farcs{}6 degrades $\sim$ 20--50\% more than the SPC's on average, up to a factor of two worse (at $\rho$ $\sim$ 0\farcs{}35 and 0\farcs{}55).  

\begin{deluxetable*}{llllcccccccc}
\setlength{\tabcolsep}{0pt}
\tablecolumns{8}
\tablecaption{SCExAO Observations}
\tiny
\tablehead{{UT Date} & {UT Start Time} & {Target} & {Coronagraph}& {Estimated $H$-band Strehl Ratio} & {t$_{\rm int}$ (s)} & {N$_{images}$} & {Parallactic Angle Motion ($^{\circ}$)}}
\startdata
\hline
\\
2016-07-17 & 13:15:49	   &    HR 8799 & Vortex & 80? & 30 &  110 &   167.1\\\\
\hline
\\
2017-07-26 & 13:38:14 & HR 8799 & SPC & 70? & 20.65 & 139 & 156.4\\\\
\hline
\hline
\\
2017-09-06 & 11:21:37 &HD 1160 & Lyot &  80 & 45.7 & 3 & 1.45\\
	   & 	11:24:29 & "     &   "   & 75     &   "   & " & 2.23\\
   	   & 11:28:08 &	"     &   "   & 70     &   "   & 4 & 1.50\\
\\
	   &  11:32:47 &  "     & SPC   & 70     & 45.7  & 3 & 1.52\\
	   &   11:35:37 & "     & "   & 75     & "  & 4 & 1.51\\
	   &  11:39:06 &  "     & "   & 80     & "  & 3 & 2.28
 \enddata
\label{onskydata}
\end{deluxetable*}

\section{On-Sky Testing and Validation of the Shaped Pupil Coronagraph}
While our internal source experiments testing the shaped-pupil coronagraph reaffirm the design's promise, on-sky observations of known directly-imaged substellar companions provide a more decisive assessment.    Our experiments provide a reasonable model of the influence of low-order aberrations on the degradation of the residual halo intensity.  But on-sky data better measure the cumulative effect of both a brighter residual halo and more poorly suppressed pinned speckles.   As the SPC we use has a substantially poorer throughput than the Lyot coronagraph, a sequence of data taken with the SPC with a cumulative integration time typical for science observations provides an assessment of our SPC's suitability for exoplanet discovery and characterization right now.
Thus, we obtained on-sky data with SCExAO coupled with the SPC for two well-known stars with directly-imaged substellar companions: the A0 star HD 1160, which hosts a bright 80--90 $M_{\rm J}$ substellar companion at $\rho$ $\sim$ 0\farcs{}8 \citep{Nielsen2013,Garcia2017}, and the A5 star HR 8799, which hosts four fainter planets with masses of 5--7 $M_{\rm J}$ at $\rho$ $\sim$ 0\farcs{}39--1\farcs{}72 \citep{Marois2008,Marois2010a,Currie2011a}.     

The modest-contrast brown dwarf/low-mass star HD 1160 B has been detected already in short exposures with SCExAO coupled with the HiCIAO camera \citep{Garcia2017}.   Our new observations with SCExAO/CHARIS provide an empirically-driven assessment of the SPC's robustness in the face of low-order aberrations compared to the Lyot, complementing our experiments with the SCExAO internal source.   
Previous SCExAO/HiCIAO observations of HR 8799 with the vortex coronagraph detect all four planets at S/N $\sim$ 10 or greater \citep{Currie2017b}, providing a reference point for our attempt to detect the inner three planets using the SPC\footnote{HR 8799 b lies outside of the field of view on CHARIS.}.    Our SCExAO observations of HD 1160 and HR 8799 are summarized in Table \ref{onskydata}.

\subsection{SCExAO Shaped-Pupil and Lyot Data for HD 1160}
\begin{figure*}[t!]
\centering
\includegraphics[scale=0.35,clip]{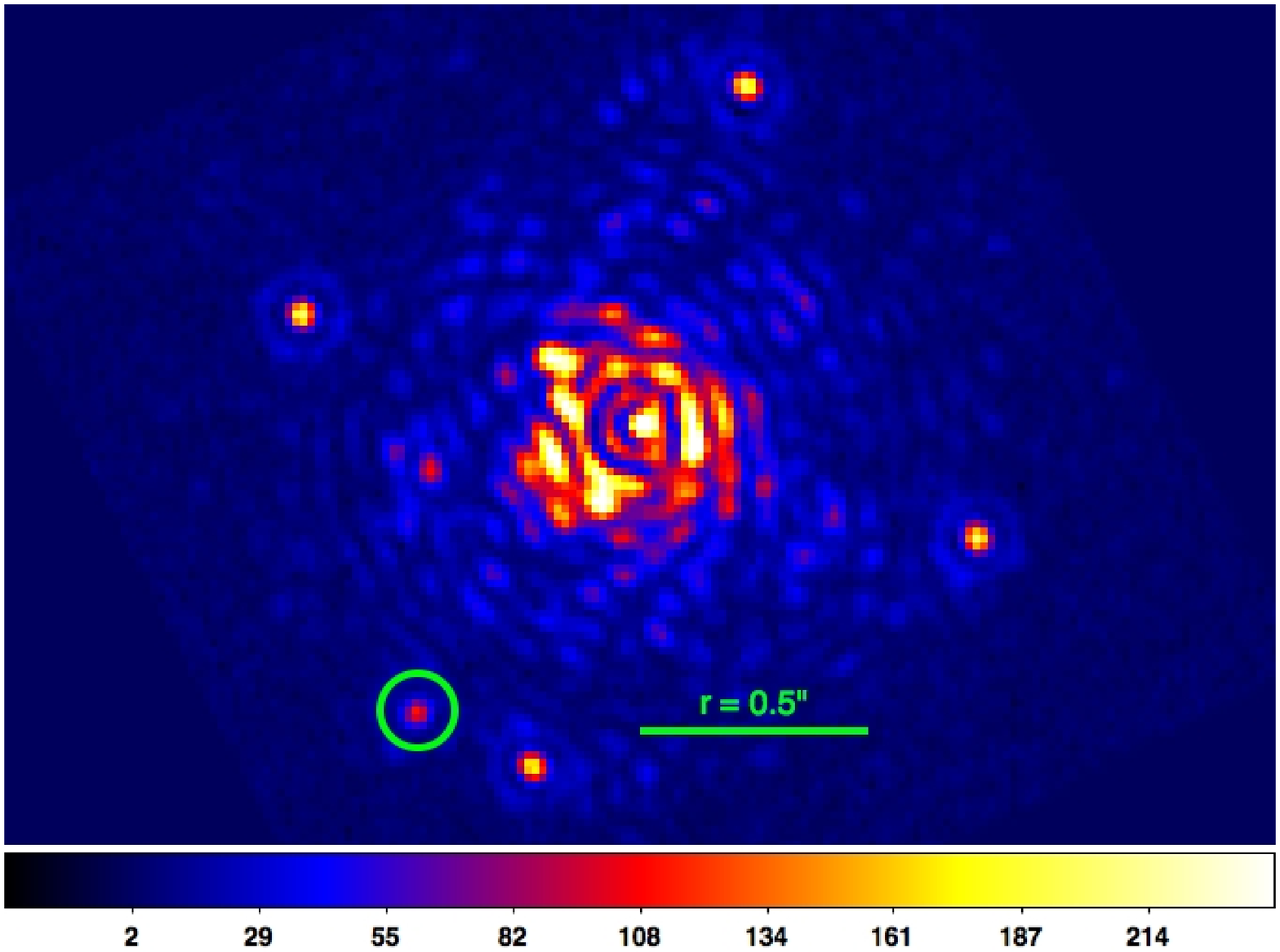}
\includegraphics[scale=0.35,clip]{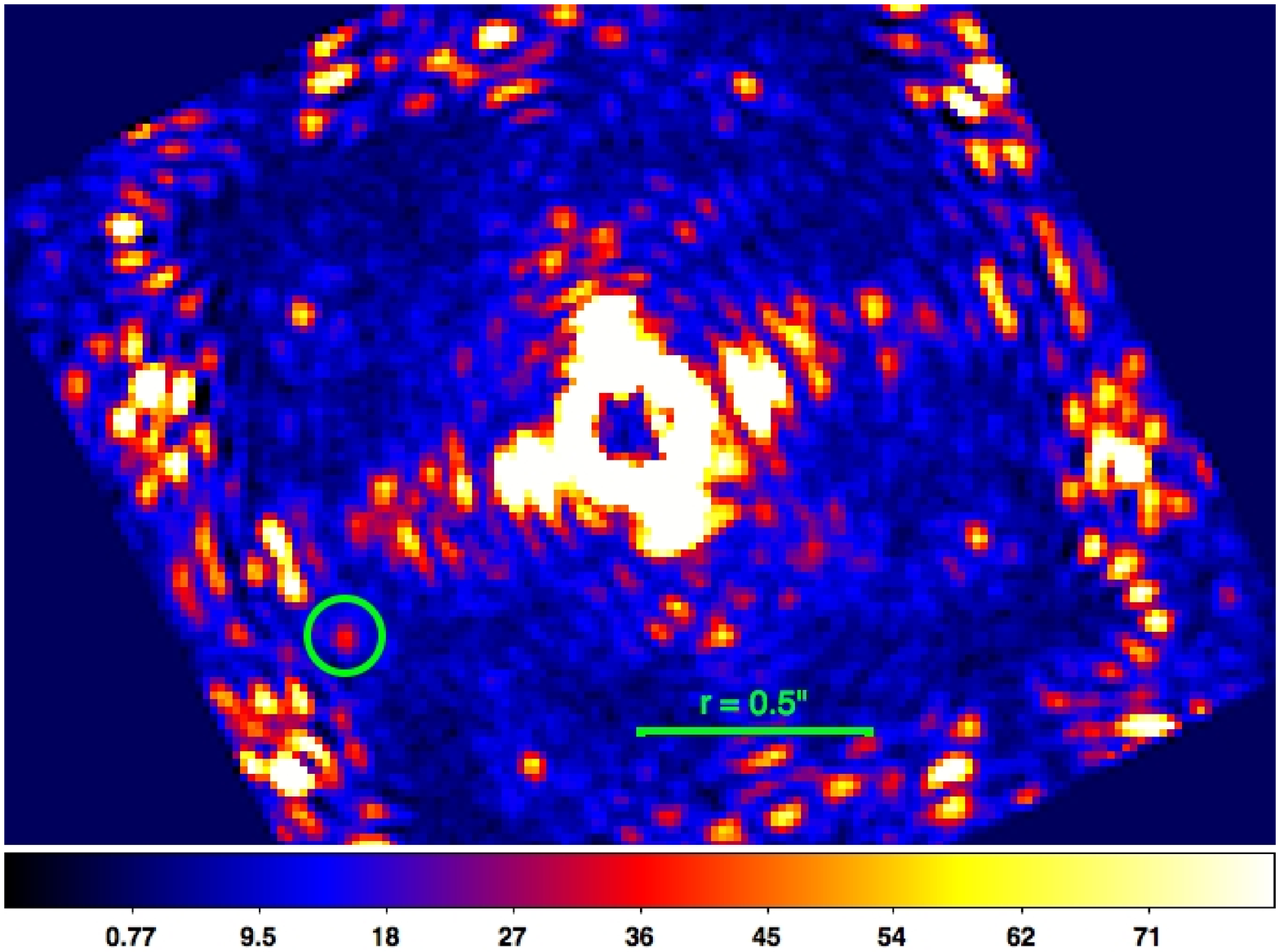}
\vspace{-0.035in}
\caption{Representative $H$-band image slice from SCExAO/CHARIS data for HD 1160 using the Lyot coronagraph (left) and the shaped-pupil coronagraph (right).   In both cases, the reported wavefront error implied a Strehl of $\sim$ 80\% at 1.6 $\mu m$.   The horizontal color bar corresponds to the raw counts in the image slice: the throughput of the shaped pupil is roughly a factor of 3.3 lower.  The position of HD 1160 B is circled: the difference in its position angle between image slices is due to parallactic angle motion ($\Delta$$\theta$ $\sim$ 20$^{o}$).  The companion's contrast in this image slice is roughly 10$^{-3}$.}
\label{lyotspccharis_repimage}
\end{figure*}

\subsubsection{Observations and Image Processing}
We obtained SCExAO observations of HD 1160 on 6 September 2017  with
 the CHARIS integral field spectrograph operating in low-resolution ($R$ $\sim$ 20), broadband (1.13--2.39 $\mu m$) mode \citep{Peters2012,Groff2013,Groff2015,Groff2016}.
SCExAO/CHARIS data were obtained using the shaped pupil (without a Lyot stop) and the Lyot coronagraph, both with the 217 mas diameter occulting spot.    Although the standard seeing monitor for the Maunakea summit did not report natural seeing values for this night, skies were clear with 5 mph wind and SCExAO's superb corrections on brighter stars observed before and after HD 1160 ($\sim$ 90\% Strehl in $H$ band) implied high-quality, ``slow" seeing.   Exposure times for HD 1160 consisted of 45-seconds per CHARIS data cube.

The real-time AO telemetry monitor provided both a measure of the residual wavefront error (and thus the Strehl ratio) and the contribution of this error from low-order and high-order modes organized into eleven different "blocks"  on the AO control loop.   For SCExAO's nominal correction, the inferred Strehl at 1.6 $\mu m$ was 80\%.   Block zero -- containing contributions from tip-tilt and other low-order modes (e.g. coma, focus) -- plainly dominated the AO wavefront error budget.   The dominance of tip-tilt in our error budget in particular was further confirmed by simultaneous $V$-band imaging of HD 1160 with the VAMPIRES instrument \citep{Norris2015}, revealing visible changes in the star's centroid position on $<$ 1 $s$ timescales.

To study SCExAO's performance with coronagraphs with different levels of low-order aberrations, we reduced the gain on the AO loop from $\sim$ 10\% to 6\% and then 3\%, which increased the low-order aberrations sufficiently to yield a total wavefront error corresponding to 75\% and then 70\% Strehl.   We obtained data for HD 1160 using both the Lyot and SPC at these three gain settings and thus at Strehl ratios of 80\%, 75\%, and 70\%.

The CHARIS Data Reduction Pipeline \citep[CHARIS DRP;][]{Brandt2017} converted raw CHARIS data into data cubes consisting of 22 image slices spanning wavelengths from 1.1 $\mu m$ to 2.4 $\mu m$.   Calibration data provided a wavelength solution; the robust `least squares' method described in \citeauthor{Brandt2017} extracted CHARIS data cubes with a spaxel scale of  0\farcs{}0164 and $\sim$ 1.05\arcsec{} radius field-of-view.  Reduction steps after the creation of these data cubes will be made available in a future public release of the Python-based CHARIS DRP.   For these data, we constructed a provisional data reduction pipeline written in IDL and leveraging on the code used in previous Subaru/Keck/VLT broadband imaging data \citep{Currie2011a} and Gemini/GPI integral field spectrograph data \citep{Currie2015a,Currie2015b}.   

  Briefly, we first converted the CHARIS fits header to a standard syntax and used exposures taken with CHARIS dithered well off of HD 1160's position for sky subtraction.   For each data cube, we estimated the centroid positions of the satellite spots in spatially-filtered versions of each of the 22 image slices.  To fine tune our estimates of the star's position and mitigate the effects of atmospheric dispersion, we modeled the star's position in each wavelength slice as a 3rd order polynomial with robust outlier rejection.   Spectrophotometric calibration follows previous steps taken for GPI data in \citet{Currie2015a,Currie2015b}, which themselves are slight modifications of primitives used in the GPI Data Reduction Pipeline, version 1.4.0 \citep{Perrin2014}.   Our calibration uses HD 1160's published photometry and spectral type and an A0V standard from the Pickles spectral library \citep{Pickles1998} binned to CHARIS's resolution and interpolated into its wavelength axis.   The satellite spots throughput is not a constant across each channel as with GPI but varies as $atten_{\lambda}$ = $atten(1.55 \mu m)$ $\times$($\lambda$/1.55 $\mu m$)$^{-2}$, where the attenuation is measured in linear units ($atten(1.55 \mu m)$ = 2.72$\times$10$^{-3}$). 
  
 To assess the performance of the shaped pupil and Lyot coronagraphs, we first measured the NIP using the same methods as with our internal source experiments and then investigated the stability of the halo and residual noise after point-spread function (PSF) subtraction.   We used a simplified version of the standard locally optimized least-squares algorithm \citep[LOCI][]{Lafreniere2007a} to perform PSF subtraction in large annular regions\footnote{As our sequences were very short, we have too few degrees of freedom available to justify using more complex PSF subtraction methods \citep[e.g.][]{Marois2010b,Currie2012}; the weighted reference image sections used to construct a reference PSF section by LOCI are little different from a straight median-average PSF.   Thus, only slightly poorer results should be obtained with classical PSF subtraction as in \citep{Marois2006}.   Furthermore, since these short sequences have little parallactic angle motion, ``contrast curves" do not describe the detectability of astrophysical objects.   Rather, they should simply be viewed as a measure of the stability of the PSF (as probed by the subtraction residuals) on short, 5-10 minute timescales.}.   Following previous work \citep[e.g.][]{Currie2015a,Currie2017}, we computed contrast curves  by replacing each pixel by the sum in each image slice (and broadband image) within a 1 $\lambda$/D aperture, measuring the radial profile of the robust standard deviation of this summed image, correcting for small sample statistics \citep{Mawet2014}, and dividing by the stellar flux.     
 \subsubsection{General Features of the Shaped Pupil and Lyot Data}
 Figure \ref{lyotspccharis_repimage} shows representative $H$-band image slices of CHARIS data cubes for HD 1160 obtained using the Lyot coronagraph (left panel) and shaped pupil (right panel).    The Lyot data show four satellite spots with a clear Airy ring pattern, consistent with our estimate of 80\% Strehl at 1.6 $\mu m$ for this data cube.   HD 1160 B (circled) is plainly visible at $\rho$ $\sim$ 0\farcs{8}.   The shaped pupil image slice also reveals HD 1160 B in the raw data.   The SPC's dark hole region is set at 16 $\lambda$/D for each slice, which corresponds to $\sim$ 0\farcs{}5--1\farcs{}0 from $J$ to $K_{\rm s}$ band.   Thus, it is harder to differentiate HD 1160 B in the displayed slice and at shorter wavelengths from the bright outer halo at $\rho$ $>$ 0\farcs{}85 and from the bright cross-like pattern at smaller separations.    At $K$-band, it is much better separated as the four dark regions are larger.   

Furthermore, as implied by the color bars, the throughput with the shaped pupil is far poorer than the Lyot coronagraph, in spite of not using a Lyot stop.  Compared to the Lyot coronagraph, the SPC throughput is a factor of $\sim$ 3.3 $\pm$ 0.2 worse.   Therefore, to reach the same sensitivity at wider separations, where detections become background limited, the data with this SPC must be nearly a factor of 11 deeper.

 \begin{figure*}[t!]
\centering
\includegraphics[scale=0.48,trim=10mm 5mm 5mm 10mm,clip]{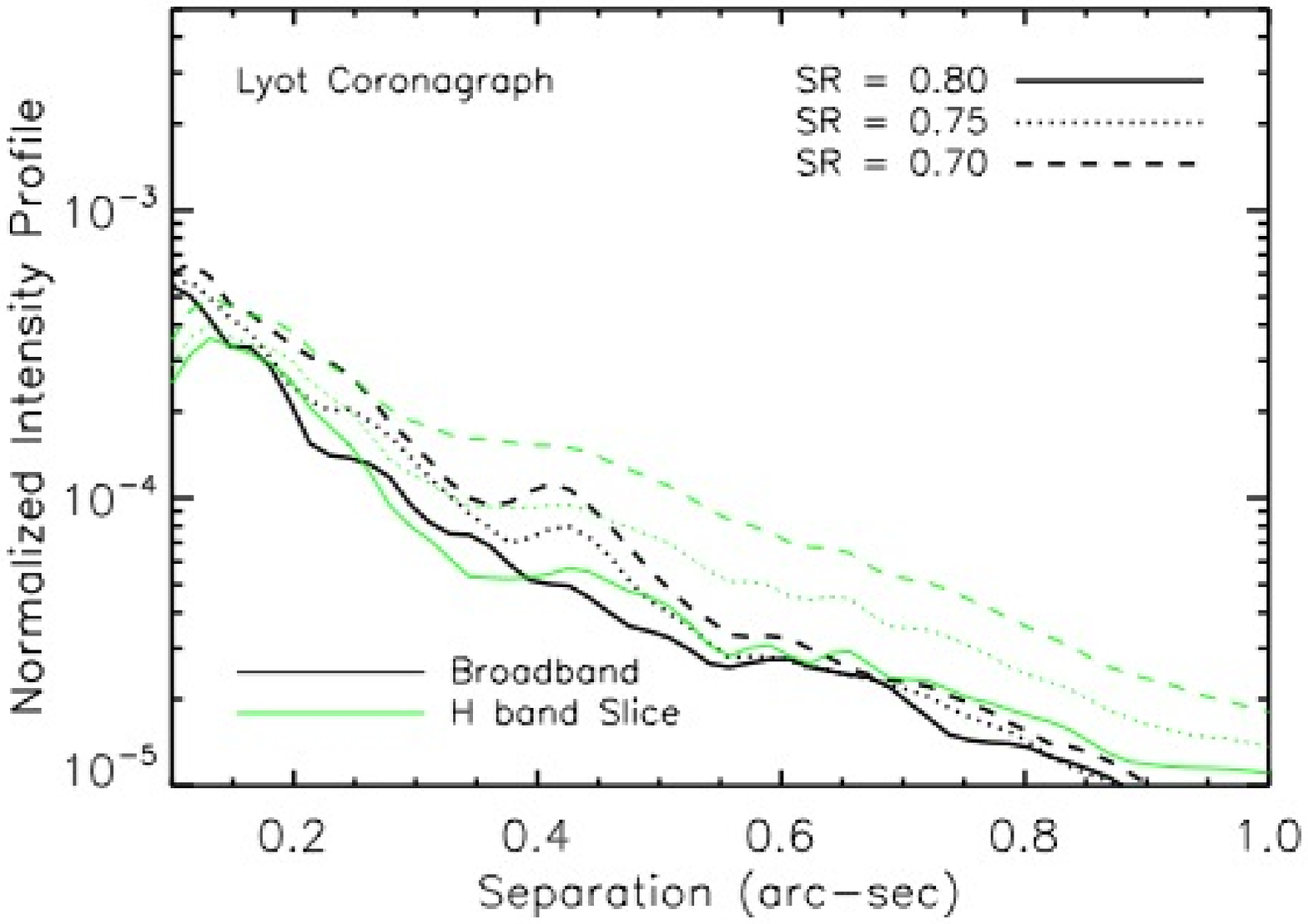}
\includegraphics[scale=0.48,trim=10mm 5mm 5mm 10mm,clip]{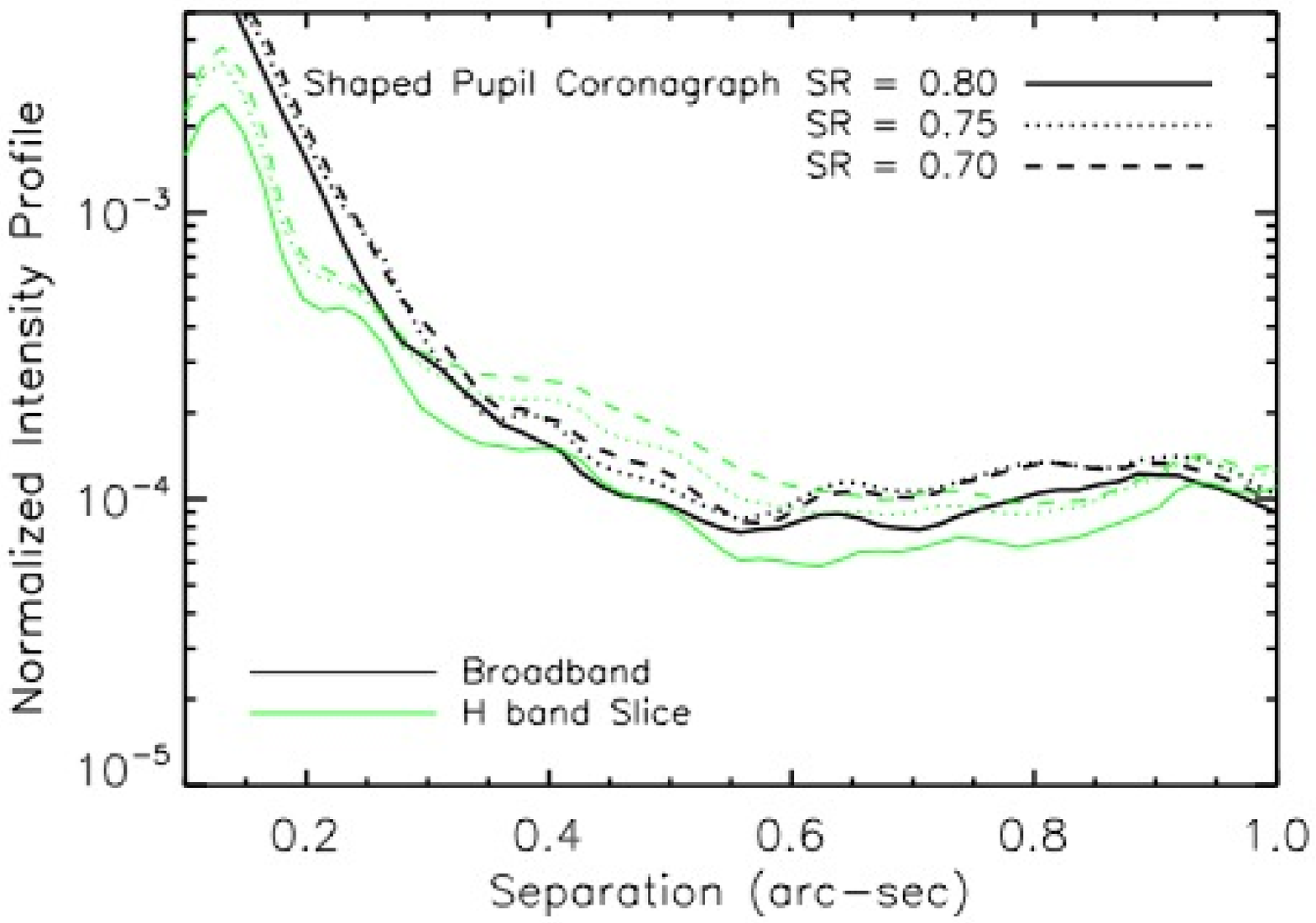}\\
\includegraphics[scale=0.48,trim=10mm 5mm 5mm 10mm,clip]{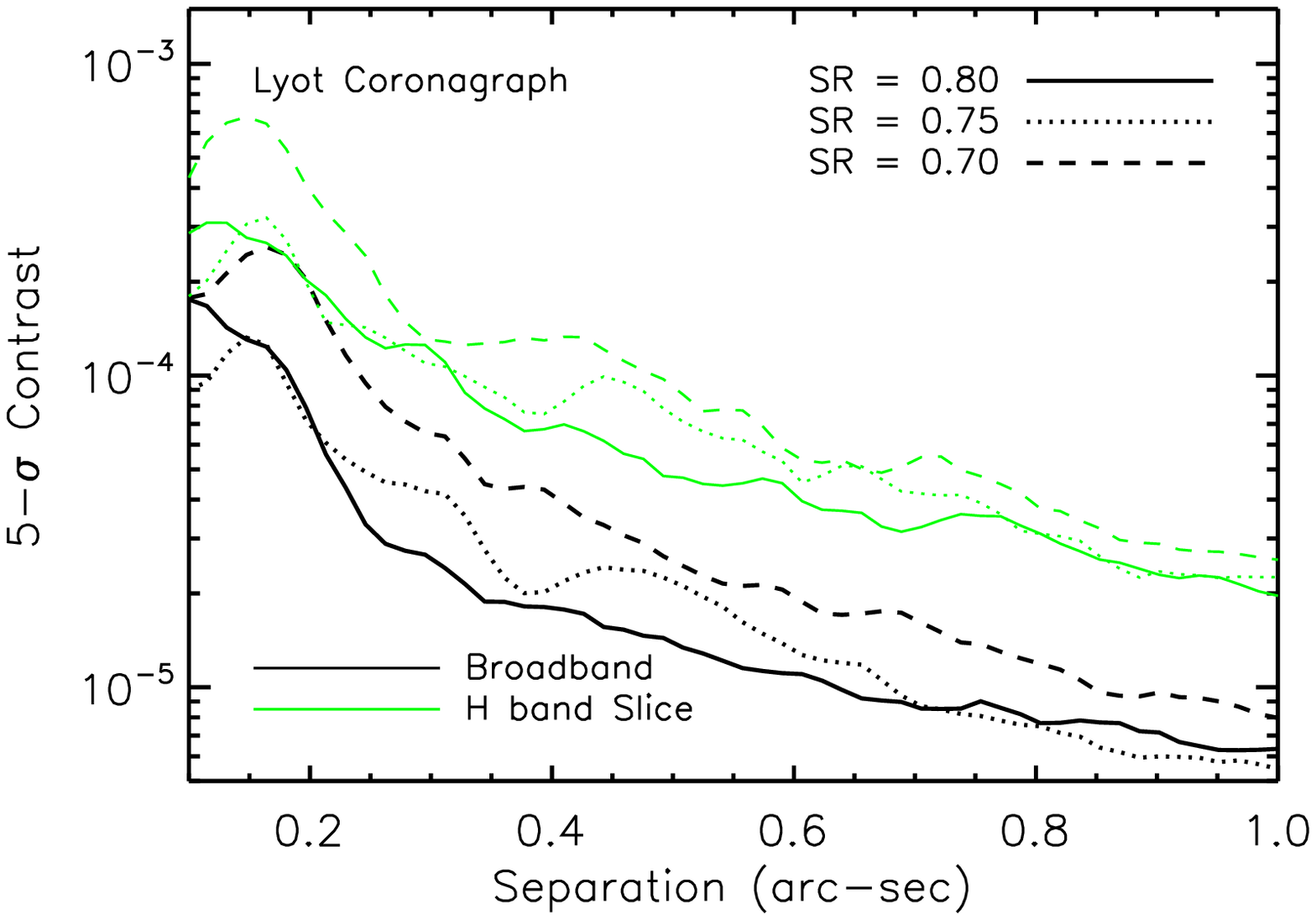}
\includegraphics[scale=0.48,trim=10mm 5mm 5mm 10mm,clip]{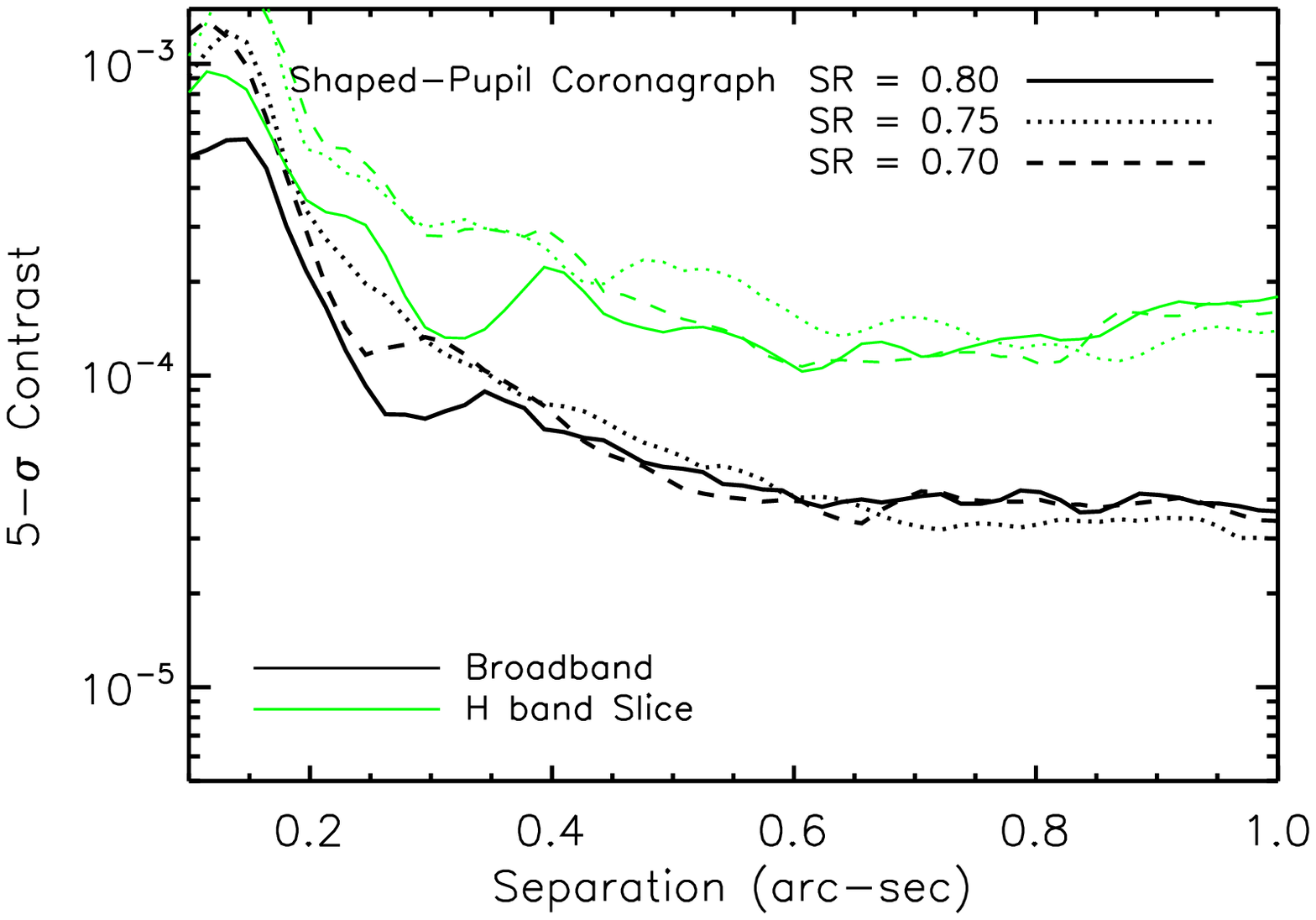}

\vspace{-0.04in}
\caption{Sequence-averaged on-sky normalized intensity profiles (top panels) and 5-$\sigma$ contrasts computed from the median-combination of PSF-subtracted data cubes (bottom panels) for Lyot coronagraph data (left panels) and shaped-pupil data (right panels) for data with Strehl Ratios of 0.70, 0.75, and 0.80.  We display the halo profiles and contrast curves for wavelength-collapsed broadband images (black) and for a representative slice in $H$-band (green). }
\label{lyotspc_onskycontrast_hd1160}
\vspace{-0.05in}
\end{figure*}

\begin{figure*}[ht!]
\centering
\includegraphics[scale=0.45,trim=0mm 0mm 0mm 0mm,clip]{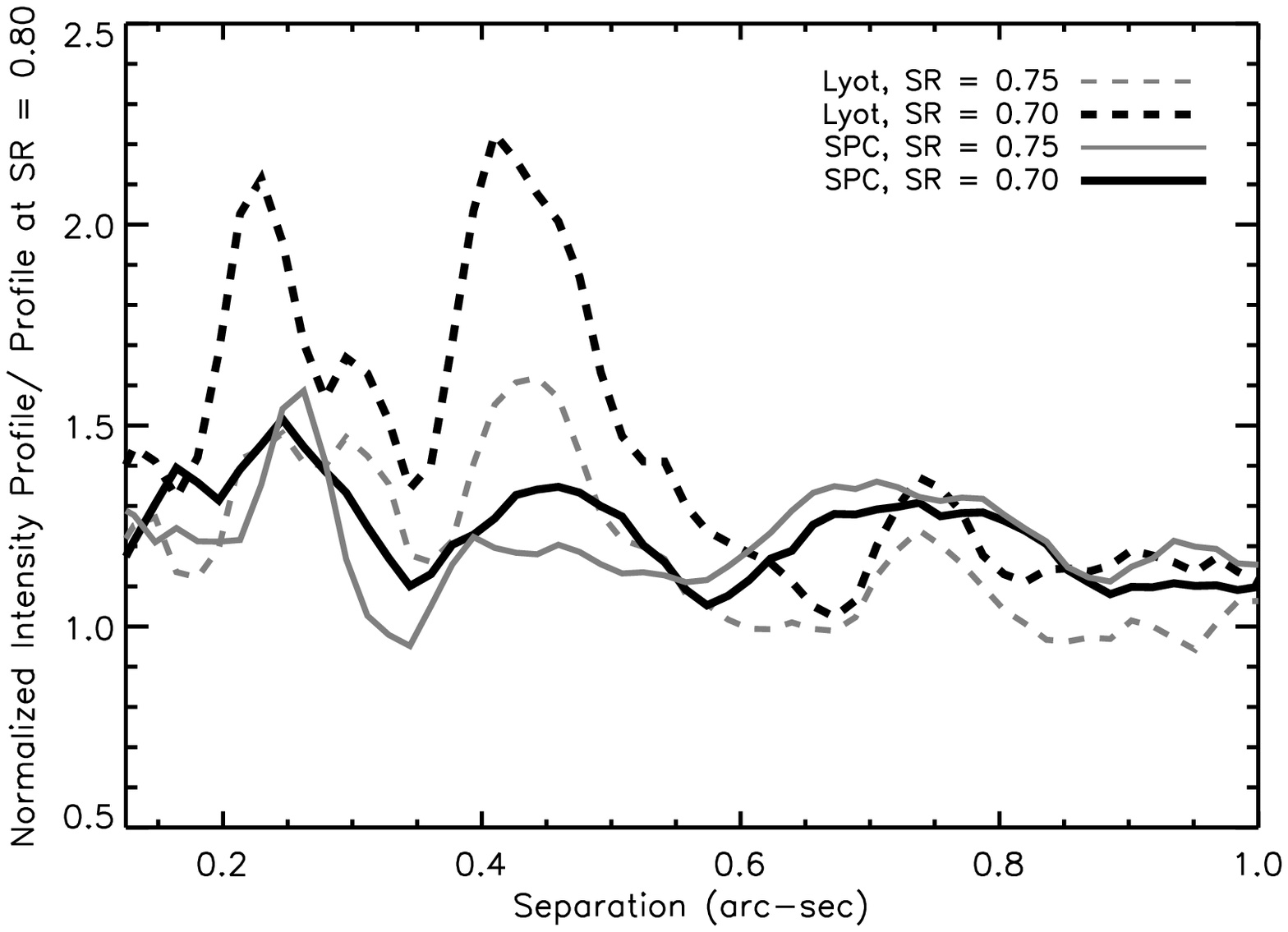}
\includegraphics[scale=0.45,trim=0mm 0mm 0mm 0mm,clip]{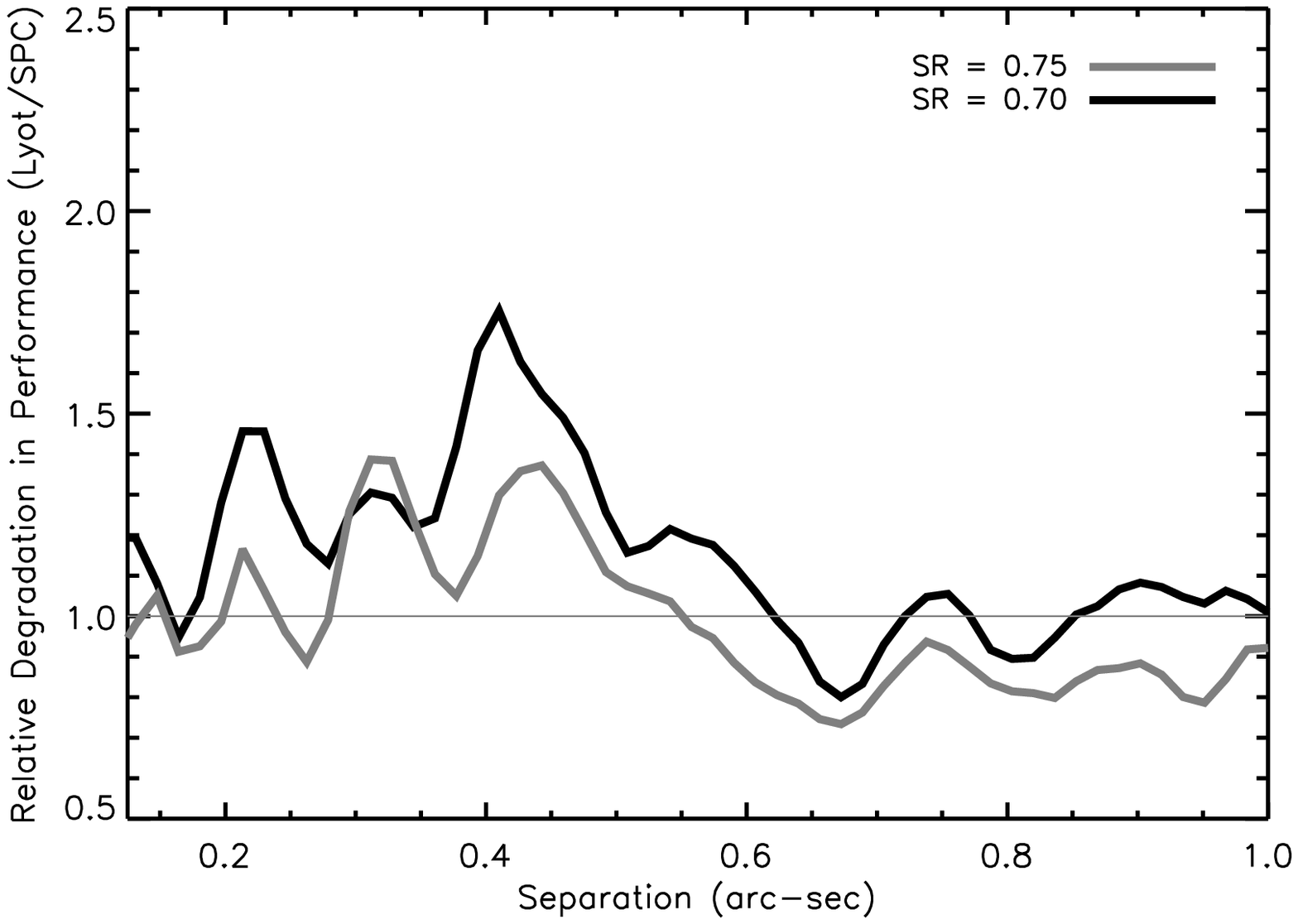}\\
\includegraphics[scale=0.45,trim=0mm 0mm 0mm 0mm,clip]{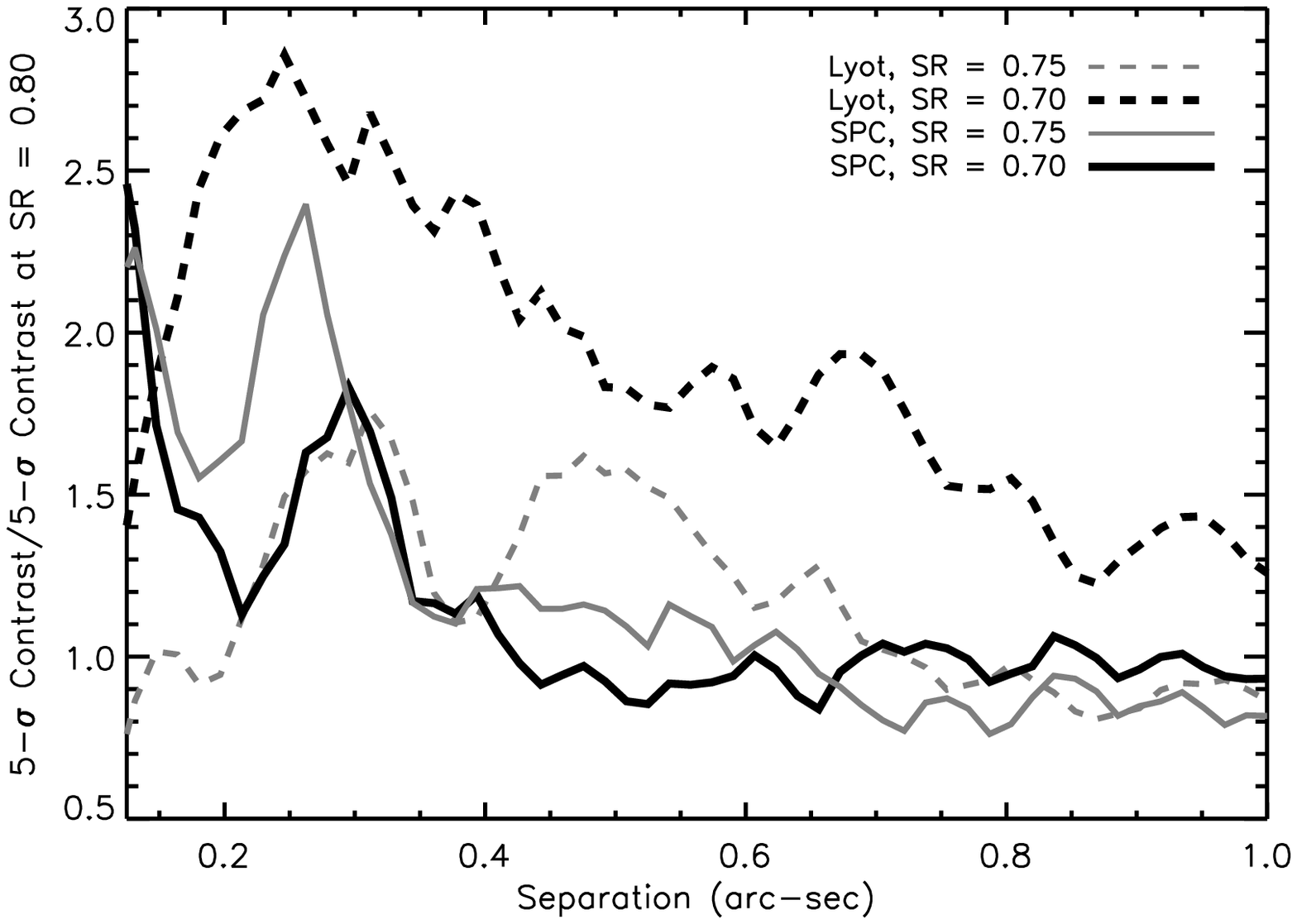}
\includegraphics[scale=0.45,trim=0mm 0mm 0mm 0mm,clip]{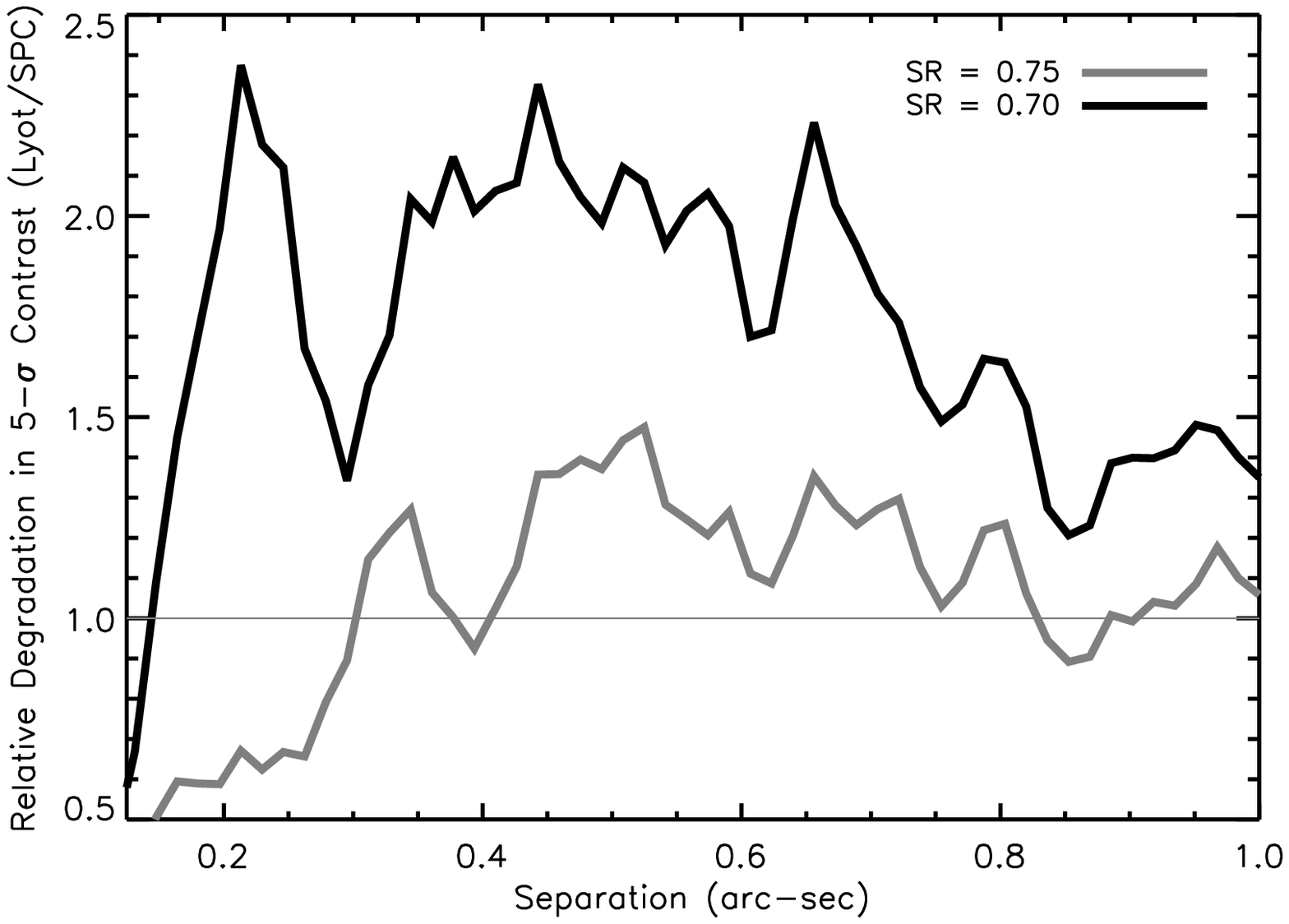}

\vspace{-0.02in}
\caption{(Top panels) On-sky analogues to Figure \ref{lyotspc_simrel}: \textit{(top-left)} The NIP at a given Strehl relative to its value at a Strehl of 0.80 for the shaped pupil and Lyot coronagraphs, and (\textit{top-right}) the RDP for the Lyot/shaped pupil.   To obtain the right-hand panel curves, divide the curves represented in the left panel for the Lyot coronagraph by those for the SPC (e.g. SR = 0.70 in the righthand panel is the lefthand panel curve labeled Lyot, SR = 0.70 divided by the curve labeled SPC, SR = 0.80).  Our on-sky data confirm that the shaped pupil's contrast is less sensitive to low-order aberrations than the Lyot coronagraph, at least for Strehl ratios below 0.75. (Bottom panels) The same plots as in the top panels except for the 5-$\sigma$ contrast, as described in \S 4.1.3.}
\label{rellyotspc_onskycontrast_hd1160}
\vspace{-0.05in}
\end{figure*}

\newpage
\subsubsection{Results}
Figure \ref{lyotspc_onskycontrast_hd1160} displays the normalized intensity profiles and 5-$\sigma$ contrast curves for data obtained with the shaped pupil and Lyot coronagraphs and at Strehl Ratios of 0.70, 0.75, and 0.80.   As seen with the internal source data, the intensity profile for the Lyot coronagraph data shows a noticeable degradation as the Strehl drops from 0.80 to 0.70 in broadband images (black curves) and in representative wavelength slices (e.g. at $H$ band, green curves), whereas the shaped pupil intensity profile is less sensitive to Strehl ratio.    The 5-$\sigma$ "contrast" likewise noticeably degrades over the same range in Strehl whereas the shaped pupil better maintains its suppressed halo.

Taking the ratio of the normalized intensity profiles and contrast curves between the Lyot and shaped pupil data at a given Strehl provides some empirical evidence that the shaped pupil's contrast is less sensitive to wavefront errors dominated by low-order aberrations (Figure \ref{rellyotspc_onskycontrast_hd1160}).  The contrast drop for Lyot data from 80\% Strehl to 70\% Strehl degrades about $\sim$ 50\% more than does the shaped pupil.   At least exterior to r $\sim$ 0\farcs{}2--0\farcs{}3, the rms of median-combined, PSF subtracted images increases a factor of 1.5--2 times more from 80\% Strehl to 70\% Strehl for the Lyot than for the shaped pupil, a difference larger than predicted based on our laboratory data (i.e. compare the curves for SR $\sim$ 0.79 to 0.57 in Figure \ref{lyotspc_simrel}).   Thus, not only do low order aberrations affect the intensity profile more strongly for Lyot coronagraphs than for the shaped pupil, but they may also lead to a more poorly subtracted halo and thus brighter contrast curve.

\begin{figure*}[ht!]
\centering
\includegraphics[scale=0.4,trim=30mm 0mm 30mm 0mm,clip]{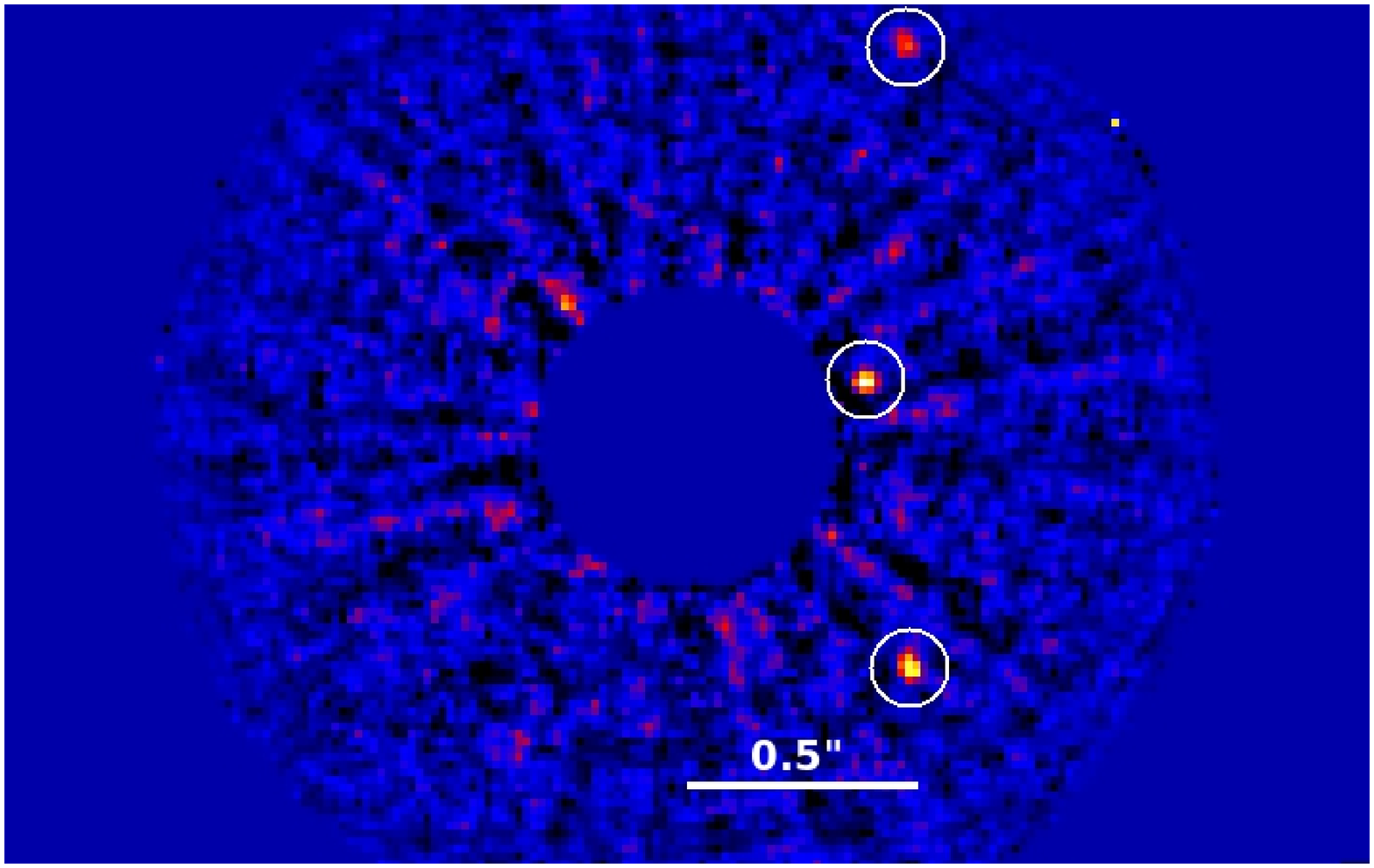}
\includegraphics[scale=0.4,trim=30mm 0mm 30mm 0mm,clip]{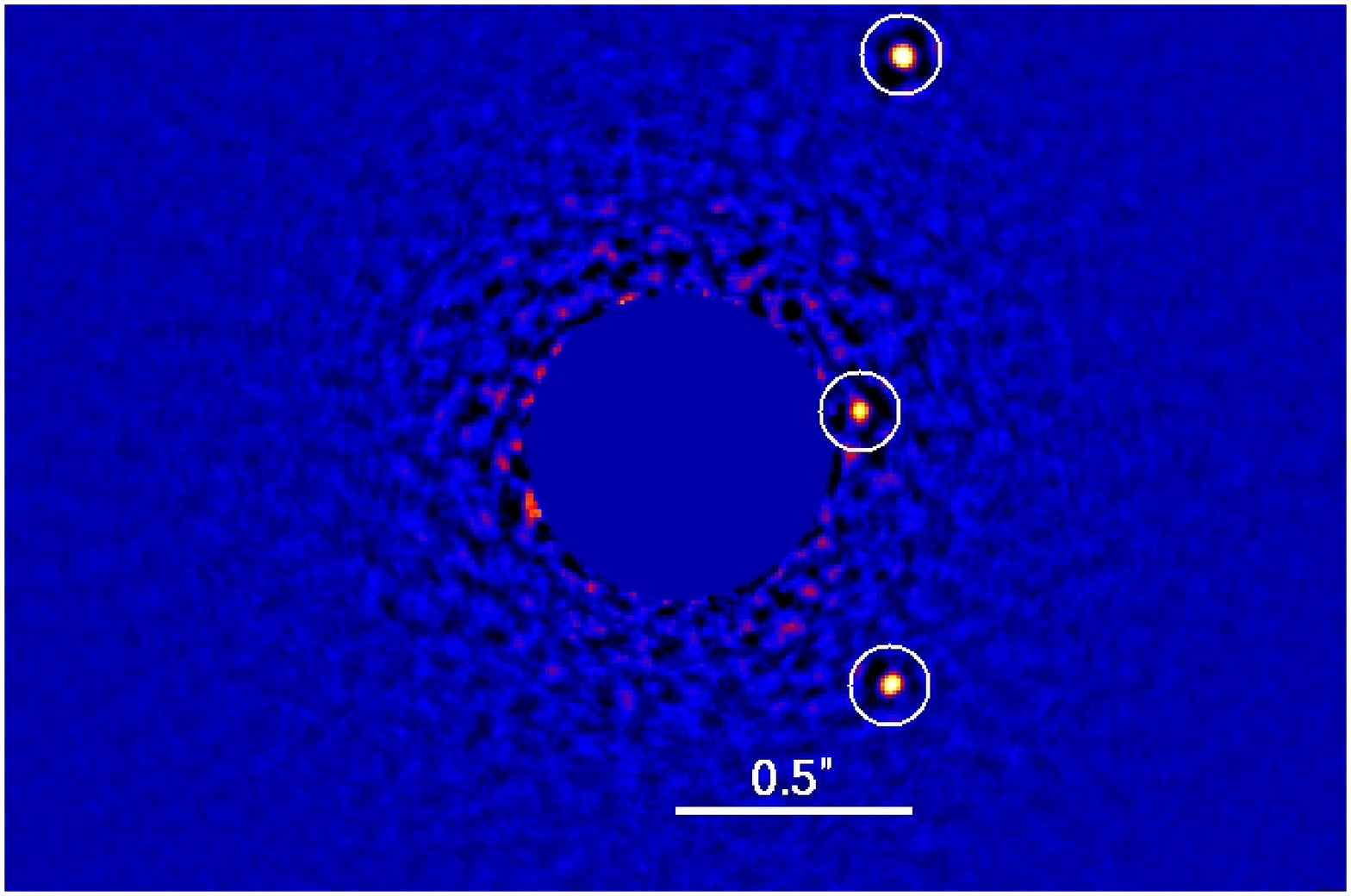}
\vspace{-0.02in}
\caption{(Left) Final wavelength-collapsed, sequence-combined SCExAO/CHARIS data for HR 8799 obtained with the shaped-pupil coronagraph.   (Right) A comparison image from 2016 obtained with SCExAO/HiCIAO in $H$ band using the vector vortex coronagraph.   Both data sets were reduced using A-LOCI with similar settings and
yield detections of HR 8799 cde; the HiCIAO data cover a wider field of view and also detect HR 8799 b (not shown).   While the vortex data yield far stronger detections of HR 8799 cd, the S/N on HR 8799 e is only slightly better than for the shaped-pupil (SNR $\sim$ 10.5 vs. $\sim$ 7).   Detections achieved with KLIP range between having 50\% lower signal-to-noise to a factor of two lower signal-to-noise versus. A-LOCI.}
\label{hr8799images}
\end{figure*}
\subsection{SCExAO Shaped-Pupil and Vector Vortex Observations of HR 8799}
On 14 July 2017, we targeted HR 8799 with SCExAO/CHARIS coupled with the shaped-pupil coronagraph and using a 2.8 $\lambda$/D radius occulting spot.   Conditions were fair for Maunakea with an optical seeing of 0\farcs{}7, low wind, and low humidity.   While we did not directly measure the Strehl ratio on HR 8799, data analyzed for other stars are consistent with a slightly below-average performance (i.e. S. R. $\sim$ 0.70 at 1.6 $\mu m$).  

The science observations consisted of 140 co-added 20.65 $s$ exposures and were obtained in \textit{angular differential imaging} \citep{Marois2006}, yielding a total parallactic angle rotation of 156$^{o}$.    As with the HD 1160 data, satellite spots provided spectrophotometric calibration and precise image registration for each data cube.   The intensity profiles for raw HR 8799 data interior to $\rho$ $\sim$ 0\farcs{}4 appear roughly the same as the HD 1160 data at 70--75\% Strehl.   Although HR 8799 is significantly brighter than HD 1160, our minimum contrast at wider separations is not any better, likely because we were unable to obtain good empty sky frames.   We constructed data cubes and performed image registration using methods described in the previous section.  For PSF subtraction, we considered both PCA and A-LOCI \citep{Soummer2012,Currie2012}, opting for the latter due to its better suppression of speckle noise.     

The SCExAO/HiCIAO 2016 H band data obtained with the vector vortex coronagraph first presented in \citet{Currie2017b} provide a comparison to our SPC data.  Briefly, methods for basic image processing and registration drew from \citet{Currie2011a,Currie2017}.   PSF subtraction follows the same approach with A-LOCI used to reduce the HR 8799 shaped-pupil data.

Figure \ref{hr8799images} displays the wavelength-collapsed, median-combined HR 8799 CHARIS/SPC image (left) compared to the 2016 SCExAO/HiCIAO $H$-band image obtained with the vector vortex coronagraph on the right.   The vortex data were obtained in slightly better conditions than the shaped pupil data and processed using nearly identical methods, recovering HR 8799 c, d, and e at S/N $\sim$ 70, 25, and 10.5.   In comparison, the SPC yields S/N $\sim$ 9, 11, and 7 for planets b, c, and e.  Thus, while the low throughput of the SPC clearly degrades its contrast exterior to 0\farcs{}4--0\farcs{}5, the difference is significantly less in more speckle dominated regions.  

\section{Discussion}
This study provides an early, detailed evaluation of the performance of the shaped-pupil coronagraph behind a well-corrected wavefront in the presence of low-order aberrations using the SCExAO system.   With the SCExAO internal source and simulator, we explored the SPC's halo suppression vs. that of a standard focal-plane (Lyot) coronagraph as low-order aberrations increase.  The SPC's absolute contrast floor was poorer than that of the Lyot coronagraph;
however, its contrast degrades slower than the Lyot coronagraph as low-order aberrations increase.  

Our on-sky SCExAO data obtained with the SPC provide a first direct probe and quantitative assessment of the coronagraph's weaker sensitivity to low-order aberrations.   In short sequences on HD 1160 covering an $H$-band Strehl range of 70--80\%, the SPC's halo suppression and contrast degrades more slowly than the Lyot coronagraph.   The difference between the SPC and Lyot in how much the halo degrades as low-order aberrations increase may be larger than predicted from our internal source simulations alone.   In deeper sequences on HR 8799, the vortex data yield far stronger detections of HR 8799 cd than the SPC but is less advantageous in detecting the innermost HR 8799 planet (HR 8799 e)\footnote{We caution that this is study is not an assessment of the SPC's absolute contrast gain behind a real system.   While our HR 8799 sequences with the SPC and the vortex had a comparable setup and duration, they were obtained at different epochs.   Our on-sky HD 1160 sequences directly comparing the SPC and the Lyot on the same night were short in duration.   Moreover, while we consider our basic qualitative results regarding the promise of the SPC to be robust, a redesigned SPC eliminating some stellar halo leakage terms (e.g. due to manufacturing errors) may yield a slightly different mapping between intensity profile or contrast and Strehl ratio that may also imply a slightly different quantitative assessment of the coronagraph's sensitivity to low-order aberrations vs. other designs.}.

The key performance limitations with the SPC utilized in this work are the 1) lower throughput of the coronagraph and 2) shallow contrast floor with respect to what is achievable with other coronagraphs used behind SCExAO.    Both of these aspects can be improved, as the mask itself was designed to be a rapid, inexpensive demonstration of the shaped-pupil concept tailored to SCExAO's early performance, not a workhorse science instrument for a fully-commissioned SCExAO.   Revisiting the experiments described in this work using an SPC with a higher throughput and deeper contrast floor matching or exceeding that of other advanced coronagraphs such as the Phase Induced Apodization Complex Mask Coronagraph \citep[PIAACMC][]{Guyon2014} will also better assess the coronagraph's sensitivity to low-order aberrations for future high-contrast imaging platforms.

 First, providing the coronagraph with an anti-reflective coating will help increase the total system throughput.   The shaped-pupil design necessarily results in a reduction in throughput vs. a standard Lyot coronagraph.   However, with a new coating, better observing strategies (including sky frames), and improvements in the CHARIS DRPs ability to suppress instrumental noise we may provide significantly better performance at wider separations ($\rho$ $\gtrsim$ 0\farcs{}5--0\farcs{}75) where detections are background limited.

Furthermore, redesigning the coronagraph itself could lead to a deeper dark hole and contrasts competitive with the vector vortex coronagraph and the upcoming PIAACMC design, at least over smaller subsections of the image plane.   A more precisely designed mask well matched to the SCExAO pupil (including dead actuators on the system's deformable mirror) could reduce scattered light in the dark regions and close the performance gap overall.   

Finally, a different transmission profile for the SPC on Subaru/SCExAO could trade a smaller dark region for deeper contrast (e.g. see Carlotti et al. 2012).  Such a design may make the SPC poorly suited for a general direct imaging search, which benefits from access to a discovery zone comprising a large fraction of the full 360$^{o}$ coverage on the science camera/IFS.   However, the SPC could work as an architecture used for demonstrating deep contrast within well-defined regions of the image plane and follow-up characterization, rather than discovery, of already-known exoplanets using SCExAO or future extreme AO systems on 10-30 m class telescopes.   This particular focus, technology demonstration and spectroscopic characterization, is the same vision for the SPC's use with WFIRST-CGI.

\acknowledgements 
We thank the anonymous referee for helpful manuscript comments.  The development of SCExAO was supported by the JSPS (Grant-in-Aid for Research \#23340051,\#26220704 \#23103002), the Astrobiology Center (ABC) of the National Institutes of Natural Sciences, Japan, the Mt Cuba Foundation and the directors contingency fund at Subaru Telescope.
CHARIS was built at Princeton University under a Grant-in-Aid for Scientific Research on Innovative Areas from MEXT of the Japanese government (\# 23103002).
We wish to emphasize the pivotal cultural role and reverence that the summit of Maunakea has always had within the Hawaiian community.  We are most fortunate to have the privilege to conduct scientific observations from this mountain.  

{}

\end{document}